\documentclass[aps,twocolumn,prx,superscriptaddress,longbibliography]{revtex4-2}
\usepackage{graphicx}
\usepackage{dcolumn}
\usepackage{bm}
\usepackage[T1]{fontenc}
\usepackage[latin9]{inputenc}
\usepackage{textcomp}
\usepackage{amsmath}
\usepackage{mathrsfs}
\usepackage{amssymb}
\usepackage{units}
\usepackage{ulem}
\usepackage{xcolor}
\usepackage[colorlinks,bookmarks=false,citecolor=blue,linkcolor=blue,urlcolor=blue]{hyperref}
\usepackage{soul}
\usepackage{comment}

\begin{document}
\title{Photon avalanche triggered by a single photon in a bistable nonlinear optical cavity}
\author{Asian Selvakumaran}
\affiliation{Universite Grenoble Alpes, CNRS, LPMMC, 38000 Grenoble, France}
\author{Anna Minguzzi}
\affiliation{Universite Grenoble Alpes, CNRS, LPMMC, 38000 Grenoble, France}
\author{Iacopo Carusotto}
\affiliation{Pitaevskii BEC Center, CNR-INO and Dipartimento di Fisica, Universita di Trento, I-38123 Trento, Italy}
\author{Maxime Richard}
\affiliation{MajuLab, CNRS-UCA-SU-NUS-NTU International Joint Research Unit, 117543 Singapore, Singapore}

\begin{abstract}
We theoretically investigate the response of a coherently-driven nonlinear optical cavity to an additional incident single photon. 
Using a quantum description of the nonlinear dynamics that fully accounts for the quantum fluctuations of the cavity field and for the discrete nature of the incident photon, we characterize the quantum dynamics of single-photon-stimulated jumps from the low-photon-number to the high-photon-number state of the optical bistability loop.
We find that the system can exhibit a giant response to this single quantum of excitation, rooted in the phase-transition picture of optical bistability. In addition to shedding light on the role of quantum fluctuations in the non-equilibrium dynamics of a nonlinear optical cavity, our results suggest a strategy for an all-optical single-photon avalanche detector.
\end{abstract}

\maketitle

\section{Introduction}

\begin{figure}[t]
\includegraphics[width=0.9\columnwidth]{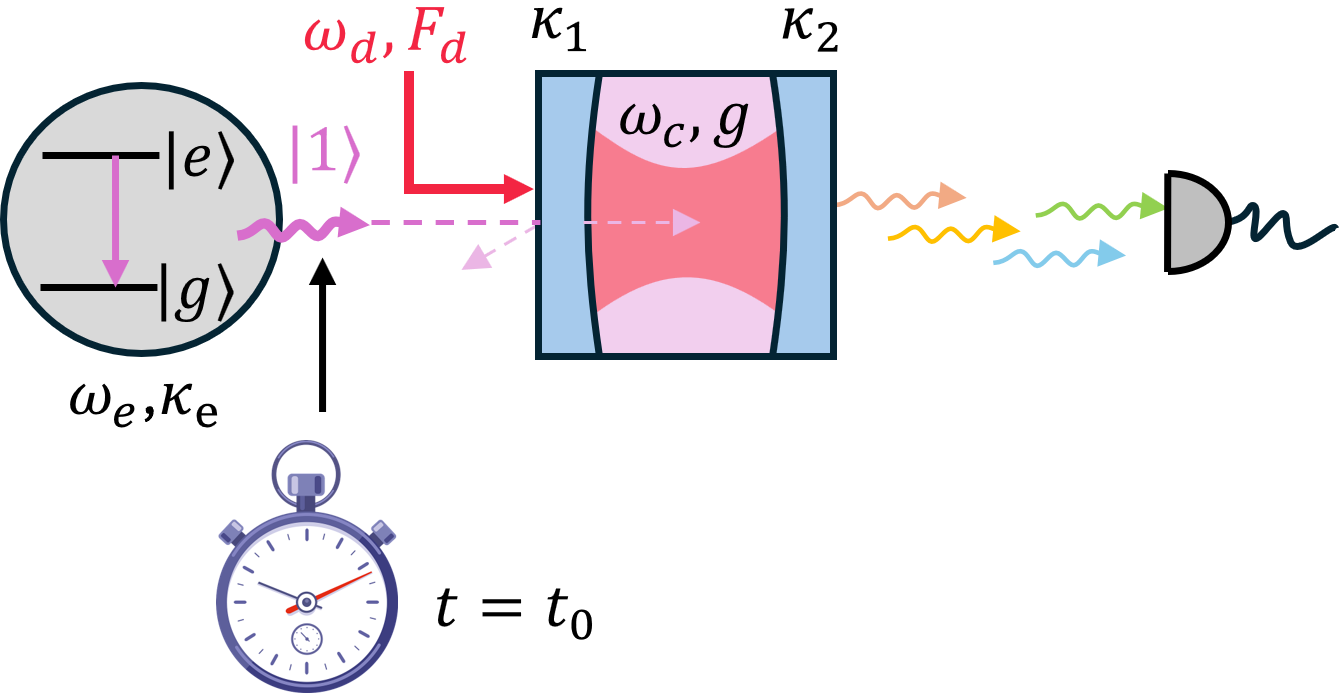}
\caption{Sketch of the system under study. We consider a single mode cavity of bare resonance frequency $\omega_c$, amplitude decay rate $\kappa=\kappa_{1}+\kappa_{2}$ resulting from the finite transmittivity of the front and the back mirrors, and Kerr optical nonlinearity $g$. The cavity is coherently driven by a monochromatic field of amplitude $F_d$ and frequency $\omega_d$. The system is initially prepared in the lower state of the bistability loop. At $t=t_0$, a single photon is emitted by a two-level system (left gray circle) of frequency $\omega_e$ and radiative rate $\kappa_{e}$ and hits the left mirror of the cavity. The quantum dynamics of the cavity field in response to the incident single photon is monitored via the intensity of the transmitted coherent drive beam (wavy arrows). }
\label{fig:fig1}
\end{figure}

When a physical system is biased into a critical regime, i.e. within a parameter region close to a phase transition point, it typically exhibits a strongly reinforced response to external perturbations: a slight perturbation can steer the system across the phase transition into a macroscopically different state~\cite{huang2008statistical}. Very soon, this striking behavior has been recognized as providing a powerful strategy to realize ultrasensitive detectors. Bubble chambers, for instance, are based on a superheated liquid that vaporizes along the path of a single charged particle traveling through it, thus revealing the particle's trajectory as a string of tiny bubbles~\cite{Wilson_1911,Glaser_1955}. Another example are superconducting nanowire biased close to the critical current: when a single photon is absorbed by the meandering nanowire track, it creates a hot spot that steers locally the superconductor into the normal state, and hence generates a measurable voltage pulse across the device; such a system constitutes the current state-of-the art for visible to infrared single photon detection~\cite{Irwin_1995,Natarajan_2012}. 

As well-established by a vast body of literature~\cite{walls2008quantum,Carusotto_2013,bloch2022non}, phase transitions can also occur in driven-dissipative photonic systems. A most celebrated example is the laser threshold, which has been described as the critical point of a non-equilibrium phase transition that spontaneously breaks the continuous $U(1)$ symmetry associated to the phase of the optical field~\cite{graham1970laserlight,Degiorgio_1970}. Similarly, degenerate parametric oscillators crossing the oscillation threshold can be described in terms of a phase transition that spontaneously breaks a discrete $Z_2$ symmetry~\cite{Gatti_1995}. Nonlinear optical cavities  under laser illumination can also exhibit a rich phase diagram when they are coherently driven via one-photon~\cite{Bonifacio_1978,Drummond_1980} or two-photon~\cite{Ciuti_1and2,Beaulieu_2025} pumps, with the transition between different steady-states becoming sharp in suitable limits~\cite{Ciuti_1pdrive}. Finally, superradiant phase transitions have been investigated in strongly light-matter coupled Rabi models~\cite{Bakemeier_2012,Ricardo_2017} and a Mott insulator to superfluid transition was predicted in many-body photon-blockaded systems~\cite{Greentree_2006,Angelakis:PRA2007,Hartmann:NatPhys2006,lebreuilly2017stabilizing,biella2017phase,Carmichael_2015} and
observed in~\cite{ma2019dissipatively}.

These remarkable advances in the study of phase transitions in optical systems suggest that these platforms can be promising candidates to realize phase-transition-enhanced ultrasensitive all-optical sensors. Capitalizing on pioneering works on bifurcation amplifiers based on classical nonlinear systems~\cite{Wiesenfeld:PRL1985,Vijay_2009} and on enhanced quantum parameter estimation at quantum phase transitions~\cite{Zanardi:PRA2008}, this idea has been explored for optical phase transitions of different nature. Sensing devices exploiting the diverging susceptibility to a classical external control parameter at a second-order critical point were investigated for a superradiant transition in quantum Rabi models~\cite{Garbe:PRL2020}, the parametric oscillation threshold~\cite{Roy_2021,DiCandia:NPJQI2023,Beaulieu:PRXQ2025}, and cavity opto-mechanical systems~\cite{Tang:PRA2023}. First-order transitions realizing photonic analogs of bubble chambers were instead investigated for optical parametric oscillators operated deep in the symmetry-broken phase: here, a weak incident light is detected by the jump it triggers from a metastable dark state to a macroscopic parametric emission state~\cite{Petrovnin:PRXQ2024}. Similar strategies can be implemented also in comparatively simpler systems like resonantly-driven nonlinear optical cavities. In suitable operating regimes, such systems display a bistable behavior with two distinct steady-states with very different light intensities. Such a behaviour can be understood as resulting from a dynamical phase transition \cite{Bonifacio_1978,Drummond_1980,Ciuti_1pdrive} and has been exploited to amplify the response to some external classical perturbation when the system is operated around a bifurcation point~\cite{Vijay_2009}.

In the present work we develop a quantum theory of the transitions between the two steady-states of such a bistable regime. While such jumps can spontaneously occur on a macroscopically long timescale as a result of quantum fluctuations as originally predicted in~\cite{Drummond_1980,vogel1988quantum,vogel1989quasiprobability} and experimentally investigated in~\cite{Rodriguez_2017,Fitz_2017,Fink2018}, they can be stimulated by an additional weak incident light, for instance a single photon. The bistable system can thus for instance serve as an efficient photo-detector device. While existing theoretical treatments of this physics are based on semi-classical approximations for the cavity dynamics and/or the additional incident field~\cite{Vijay_2009,Petrovnin:PRXQ2024}, our work builds upon the cascaded quantum system approach~\cite{Carmicheal:PRL1993} to describe the response of the system to an incident single-photon at a full-quantum level and, in this way, obtain quantitatively reliable predictions for the jump rates. 

In particular, we theoretically predict that jumps from the lower to the higher branch of the bistability loop can be triggered by an incident single photon perturbation and result in macroscopic avalanche processes observable in the transmitted light. Our conclusions are supported by a complete quantitative characterization of the different regimes of the jump dynamics as a function of the operating point of the coherently-driven cavity and of the incident single-photon wavepacket parameters such as frequency and bandwidth. Our predictions are interpreted using photo-detection concepts such as quantum efficiency, avalanche of photons, and dark-count rates and confirm the potential of our system as an all-optical realization of a single-photon detector.

The article is organized as follows. In Sec.~\ref{sec:system}, we present the model and the quantum cascaded system framework we use for our calculations. In Sec.~\ref{sec:steady-state}, we characterize the steady-state properties of the cavity throughout the bistability loop, first at the mean-field level, and then including the quantum fluctuations. The initial state preparation and the analysis of the spontaneous jump dynamics are described in Sec.~\ref{sec:prep-and-jump}. In Sec.\ref{sec:jump-with-pert}, we analyze the response of the cavity to a single photon sent onto the left mirror of the cavity and we characterize the single-photon-stimulated jumps between different states of the bistability loop. In Sec.~\ref{sec:analysis}, we then quantitatively characterize the system's response in terms of spontaneous and single-photon-stimulated jump-up rates and photo-multiplication factor. This allows to identify the parameter regimes that maximize the cavity response to the single incident photon. Finally, Sec.\ref{sec:conclusions} contains our concluding remarks. Additional technical details are given in the three Appendices App.~\ref{app:ph_freq}, App.~\ref{app:res_T}, and App.~\ref{app:Pj}.

\section{The physical system and the theoretical model}
\label{sec:system}

In this work we consider a single-mode nonlinear cavity, driven by monochromatic coherent laser light. Our goal is to theoretically study the quantum dynamics of the cavity in response to an additional single-photon wavepacket incident on it. A sketch of the configuration is shown in Fig.\ref{fig:fig1}.

The conservative dynamics of the coherently-driven, single-mode nonlinear cavity is described by a Hamiltonian in the form~\cite{Drummond_1980,vogel1988quantum}
\begin{equation}
H_{c} = \hbar\omega_c a^\dagger a + \hbar ga^\dagger a^\dagger aa + (F_d e^{-i\omega_dt}a^\dagger+H.c.), \label{eq:Hcav}
\end{equation}
where the photon annihilation (creation) operators $a$ ($a^\dagger$) for the single cavity mode satisfy bosonic commutation relations
$[a,a^\dagger]=1$, $\omega_c$ is the cavity resonant frequency, $F_d$ is the amplitude of the coherent drive, assumed monochromatic at frequency $\omega_d$, and $g$ is the third-order Kerr optical non-linear constant characterizing the effective interaction strength between photons. With no loss of generality, this is assumed throughout the article to be
repulsive with $g>0$.

The cavity is coupled to external photon baths via the non-perfectly reflecting mirrors, that induce losses at a total rate $\kappa=\kappa_1+\kappa_2$ resulting from the sum of the contributions $\kappa_{1,2}$ of the left and the right mirror, as shown in Fig.\ref{fig:fig1}. Combining the conservative evolution via the Hamiltonian \eqref{eq:Hcav} and the dissipative dynamics induced by the losses, the quantum dynamics of the cavity field can be written in terms of the Lindblad master equation
\begin{equation}
    \partial_t \rho_c=-\frac{i}{\hbar}[H_c,\rho_c]+\mathcal{D}_{a,1}[{\rho_c}] + \mathcal{D}_{a,2}[{\rho_c}].
    \label{eq:master_eq_c}
\end{equation}
for the density matrix $\rho_c$,
where the dissipators describing photon losses through each $j=1,2$ mirror have the form
\begin{equation}
\mathcal{D}_{a,j}[\rho_c] = 2 \kappa_j\left(a\rho_c a^\dagger- \frac{1}{2} \{ a^\dagger a,\rho_c\} \right),\\
\end{equation}
and correspond, in a quantum trajectory formulation~\cite{breuer2002theory,carmichael1993open}, to jump operators of the form $c_j=\sqrt{2\kappa_j}\, a$. Equation (\ref{eq:master_eq_c}) admits a non-equilibrium steady-state solution $\rho_{c,ss}$ resulting from the dynamical balance of coherent drive and dissipation. The main features of this steady-state will be summarized in Sec.\ref{sec:steady-state}.

While this theoretical framework for the description of an open quantum system driven by coherent light is nowadays standard, much less straightforward is the study of the response of the system to incident non-classical light fields such as a single-photon wavepacket. In this work, we adopt the cascaded quantum systems approach proposed in~\cite{Carmicheal:PRL1993,Gardiner:PRL1993}, that provides a formalism that is at the same time rigorous and tractable.

The idea is to model the non-classical incident wavepacket by involving a light source that emits the desired state of light, and sends it in the direction of the open quantum system, as sketched in Fig.\ref{fig:fig1}. A single-photon wavepacket is naturally obtained as the product of a spontaneous emission process from the excited $|e\rangle$ to the ground $|g\rangle$ state of a two-level emitter: the photon frequency is directly controlled by the emitter frequency $\omega_e$ and the spontaneous emission rate $\kappa_e$ controls the temporal duration $\kappa_e^{-1}$ of the wavepacket.

In the cascaded quantum systems approach, one then considers a composite quantum system involving two subsystems: (i) the light source generating the desired quantum state of light, in our case a single-photon wavepacket); (ii) the open quantum system, in our case the single-mode nonlinear cavity. The emitter is modeled as a two-level system, with Hamiltonian
\begin{equation}
H_e=    \hbar\omega_e \sigma^+\sigma^- 
\label{eq:Hem}
\end{equation}
where $\sigma^\pm$ are the spin operators of the two-level emitter, $\sigma^+=|e\rangle\langle g|=(\sigma^-)^\dagger$.
The one-directional coupling of the emitter (i) towards the nonlinear cavity (ii) across the left mirror, is described by a suitably designed master equation for the full quantum system. This includes~\cite{Carmicheal:PRL1993} both a Hamiltonian coupling between the emitter and the cavity 
\begin{equation}
 H_{e,c} = i\sqrt{\kappa_e\kappa_1}({\sigma^+a - a^\dagger\sigma^-}),
\end{equation}
and a dissipative coupling stemming from the simultaneous coupling of the cavity and the emitter to the same radiative modes. This latter coupling amounts to replacing the left mirror jump operator $c_1=\sqrt{2\kappa_1}\, a$ by a combined jump operator
\begin{equation}
    C = \sqrt{2\kappa_e}\sigma^- + \sqrt{2\kappa_1}a
    \label{eq:dissipC}
\end{equation}
accounting for the interference between the light emitted by the emitter and reflected by the left mirror with the one emitted by the cavity via its left mirror.

Putting all terms together, the overall evolution of the system is then described by a master equation for the total density matrix in the form 
\begin{equation}
    \partial_t \rho=-\frac{i}{\hbar}[H_c+H_e+ H_{e,c},\rho]+\mathcal{D}_C[{\rho}] +\mathcal{D}_{a,2}[{\rho}].
    \label{eq:master_eq}
\end{equation}
where $\mathcal{D}_C[{\rho}] = C\rho C^\dagger- \{C^\dagger C ,\rho\}/2$ is the dissipator associated to the jump operator $C$.

It is interesting to note how in the quantum trajectories formulation~\cite{Carmicheal:PRL1993} the sum of the Hamiltonian coupling and of the dissipative one stemming from the jump operator \eqref{eq:dissipC} recovers  a unidirectional coupling $i\sqrt{4\kappa_e\kappa_1}\,a^\dagger \sigma^-$ describing the irreversible transfer of the single photon from the emitter towards the cavity. As a consequence of this form of the coupling and of the jump operator \eqref{eq:dissipC}, the steady-state of the full master equation \eqref{eq:master_eq} recovers the steady-state of the cavity-only master equation \eqref{eq:master_eq_c} tensor-producted with the emitter being in the ground state.

As a directly observable quantity, a great attention will be paid to the transmitted intensity of the coherent drive beam exiting from the right mirror of the cavity. A straightforward calculation within the input-output theory of optical cavities~\cite{walls2008quantum} gives that the corresponding photon flux operator is proportional to the in-cavity photon number,
\begin{equation}
\hat{\Phi}_{tr}=2\kappa_2 a^\dagger a\,.
\end{equation}

In the following,
we will make use of the Lindblad master equation \eqref{eq:master_eq} to obtain an exact description of the quantum dynamics of the nonlinear cavity in response to the single photon. Numerical solutions of the master equation in the different cases are obtained using the QuTiP package~\cite{johansson2012qutip,johansson2013qutip,qutip5}. 

\begin{figure}[t!]
\includegraphics[width=\columnwidth]{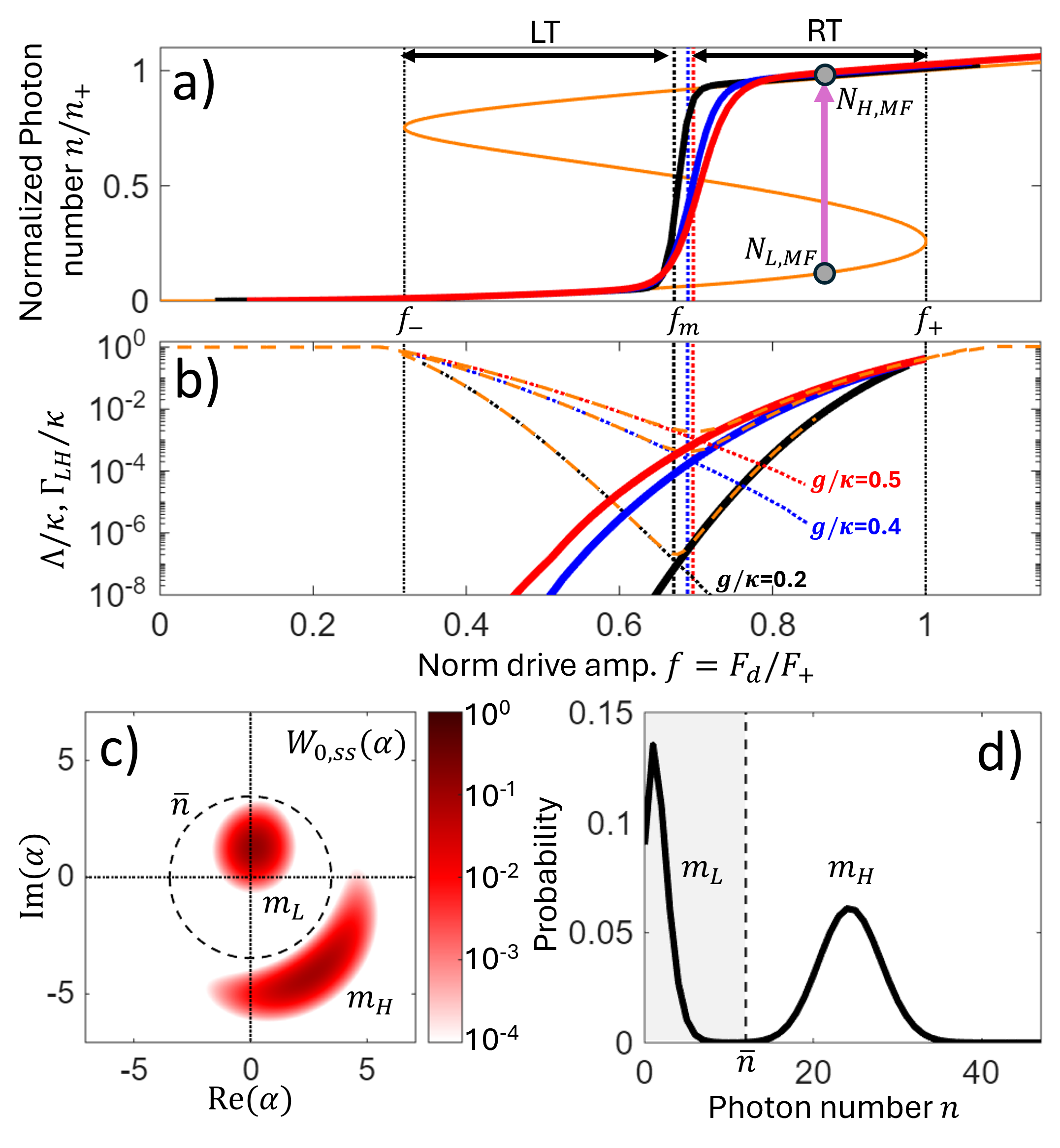}
\caption{Panel (a): normalized steady-state photon number $n/n_+$ calculated within the mean-field theory (blue) and the fully quantum theory as a function of the normalized coherent drive amplitude $f\equiv F_d/F_+$, for three different values of the nonlinearity $g/\kappa=0.2,\,0.4,\,0.5$ (black, blue, red). The plotted quantities are normalized to the coherent drive amplitude $F_+$ and the mean-field photon number $n_+$ at the right end point of the bistability loop, and $f_-$ is the same quantity at the left end point. The drive detuning is set to $\Delta\omega\equiv\omega_c-\omega_d=-8\kappa$. The lower gray circle shows an example of initially prepared $m_L$ state of mean-field photon number $n_{L,MF}$ and the upper gray point shows the corresponding $m_H$ state of mean-field photon number $n_{H,MF}$ reached via a jump-up transition (vertical arrows). Panel (b): plots of the Liouvillian gap $\Lambda$ (dashed black line), and of the spontaneous jump-up and jump-down rates $\Gamma_{LH,HL}$ (solid and dotted lines) as a function of the normalized coherent drive amplitude $f$ for $g/\kappa=0.2,\,0.4,\,0.5$ (black, blue red). $f_m=F_m/F_+$ is the normalized drive amplitude at the point where the Liouvillian gap is minimum (vertical dotted lines). The labels 'LT' and 'RT' identify the regimes where $m_{L,H}$ on the left and right side of $f_m$. Panel (c): color-plot of the Wigner representation of the steady-state intracavity field $W_{ss,0}(\alpha)$ at $f_d=f_m=0.68$ and $g=0.2\kappa$. Panel (d): corresponding intracavity photon number probability distribution $p(n)$. The dashed line in (c,d) indicates the photon number $\bar n=(n_{L,MF}+n_{n,MF})/2$ used to partition the system into the two states $m_L$ and $m_H$.
}
\label{fig:fig2}
\end{figure}

\section{properties of the steady-state in the bistable regime}
\label{sec:steady-state}

After setting up the general theoretical model, in this Section we summarize the key features of the coherently-driven nonlinear cavity dynamics in absence of perturbations. In particular, we will discuss how the bistability effects predicted by the mean-field theory get modified by the quantum fluctuations that induce spontaneous jumps between them. 

\subsection{Mean-field approximation}

Within a mean field approximation, the master equation Eq.~(\ref{eq:master_eq_c}) reduces to an evolution equation for the expectation value $\alpha=\langle a \rangle={\rm Tr } [\rho_{c} a]$ of the cavity field amplitude. Moving for convenience to the rotating frame $\tilde{\alpha}=\alpha\,e^{i\omega_d t}$, this reads
\begin{equation}
i\dot{\tilde{\alpha}}=[\Delta \omega + 2g |\tilde{\alpha}|^2 - i\kappa]\tilde{\alpha}+ F_d\,,
\end{equation}
where $\Delta\omega=\omega_c-\omega_d$ is the detuning between the empty cavity resonance and the laser frequency. The key assumption of the mean-field theory is the factorization of the interaction term, $\langle a^\dagger a a \rangle \simeq |\alpha|^2 \alpha$. 

At steady-state the field amplitude $\tilde{\alpha}$ is constant in time and satisfies 
\begin{equation}
|F_d|^2 = |{\tilde \alpha}|^2\big[\kappa^2+(2g|{\tilde \alpha}|^2+\Delta\omega)^2\big]\,.
\label{eq:MF_Steadystate}
\end{equation}
For a sufficiently negative detuning $\Delta\omega<-\sqrt{3}\kappa$, the steady-state displays bistability within an interval of coherent drive amplitudes interval bounded by the lower and upper turning points $F_-$ and $F_+$, respectively. In the following, we will normalize the drive amplitude to the upper turning point $F_+$ and use the normalized quantity $f_d=F_d/F_+$.

As it is illustrated in Fig.\ref{fig:fig2}(a), for $F_d\in[F_-,F_+]$ three steady states are possible. Two of them are dynamically stable: one with a high intra-cavity photon number $n_{H,MF}$ and another one with a low photon number $n_{L,MF}$. The third steady-state in between them is dynamically unstable. 

In agreement with the standard theory of classical optical bistability~\cite{boyd2003nonlinear}, at the level of the mean-field theory  the two steady-states are stable for indefinite times. As the system can not spontaneously jump from one steady-state to the other, one needs to steer $F_d$ across the turning points $F^\pm$ to induce a jump.

\begin{figure*}[htbp]
\includegraphics[width=\textwidth]{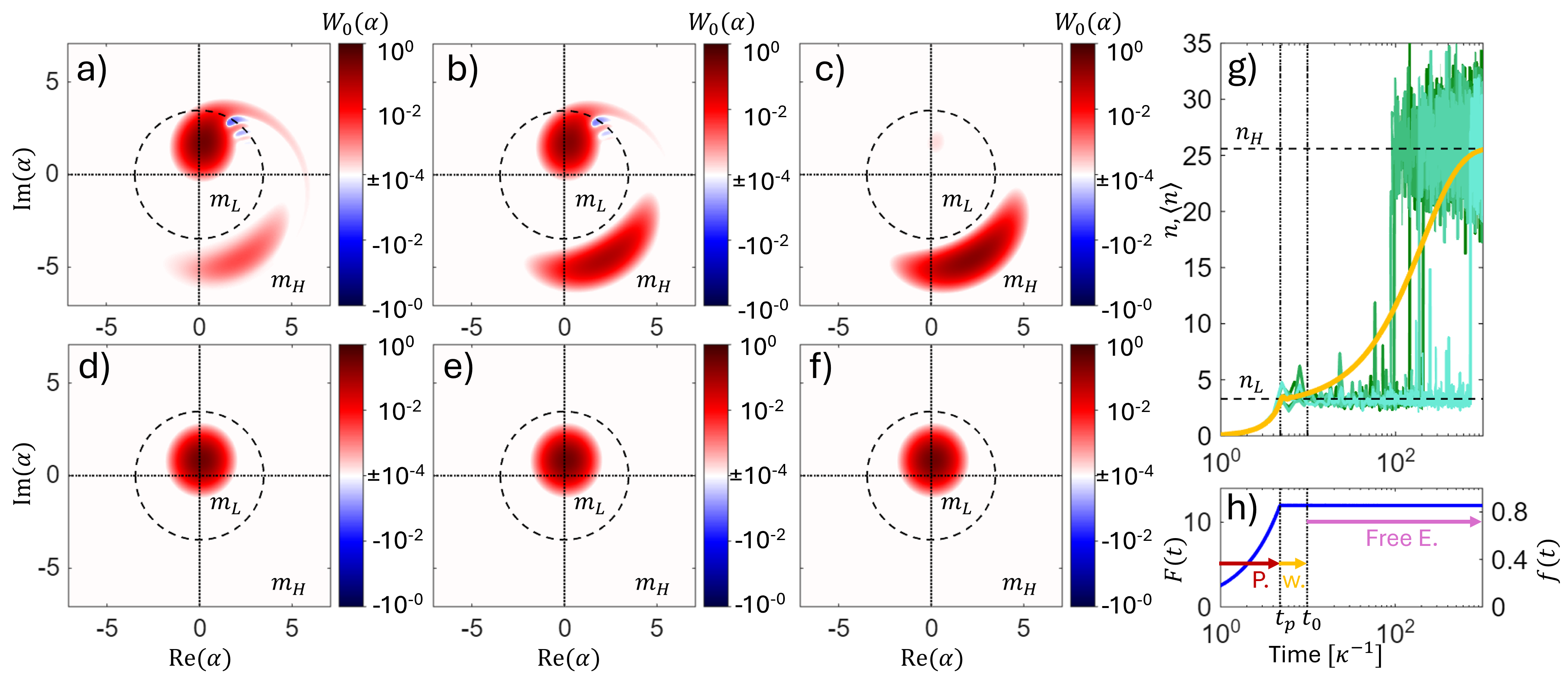}
\caption{Panels (a-f): color-plots of the Wigner function $W_0(t,\alpha)$ during the spontaneous dynamics after preparation of the system in the initial state $m_L$. In panels (a-c), the coherent drive amplitude is set at $F_d/F_+=0.856$ on the right side $RT$ of the bistable window and the snapshots are taken at $t={t_0}$ right after preparation (a), at $t=t_0+100\kappa^{-1}=t_0+0.5\Lambda^{-1}$ halfway throughout the spontaneous jump-up dynamics (b), and at $t=t_0+1500\kappa^{-1}=t_0+75\Lambda^{-1}$, a long time into the relaxation dynamics (c). In panels (d-f), the coherent drive amplitude is set at $F_d/F_+=0.461$ on the left side $LT$ of the bistable window. The snapshots are taken at $t=t_0+0.5{\kappa^{-1}}$, right after preparation (d), at later $t_c=t_0+2\kappa^{-1}$ (e) and $t=t_0+50\kappa^{-1}$ (f) times. Panel (g): plot of the time-evolution of the photon number for $F_d/F_+=0.856$ in the $RT$ region. The quantum expectation value is shown as a solid yellow line and 6 examples of individual quantum trajectories are shown in different shades of green. (h) preparation protocol (case $f=F_d/F_+=0.856$). The time-dependent normalized coherent drive amplitude is shown as a solid blue line. In the interval $t=[0,t_p]$ (labeled "P."), the drive amplitude is ramped-up from $0$ to $f$ at a rate $v_d=\kappa \partial f/\partial t$ and left constant afterwards. The residual excitations generated by the ramp-up are left to decay between $t=t_p$ and $t=t_0$ (interval labeled "W."). The perturbing photon is sent at $t=t_0$. From $t=t_0$ on, the system evolves freely ("Free E." label) under the constant drive $f$. The calculations parameters are: $\Delta\omega= -8\kappa$, $g = 0.2\kappa$, $v_d = 2.5\kappa$. In panels (a-c,g), we have used $[t_p,t_0]{\kappa}=[4.8,9.8]$. In panels (d-f), we have used $[t_p,t_0]{\kappa}=[2.58,7.58]$.}
\label{fig:fig3}
\end{figure*}

\subsection{Fully quantum theory}
\label{sec:quantumth}

When the quantum fluctuations are included in the theory, the situation changes drastically. The steady-state solution of the full master equation \eqref{eq:master_eq} is considered~\cite{Bonifacio_1978,Drummond_1980}, the curve describing the quantum-averaged intracavity photon number $n=\langle a^\dagger a\rangle$ as a function of the coherent drive amplitude $F_d$ is never hysteretic and hence does not exhibit bistability. Numerical calculations for the same parameters of the mean-field calculation are shown in Fig.\ref{fig:fig2}.(a) for $g/\kappa=0.2$ (black) $g/\kappa=0.4$ (blue) $g/\kappa=0.5$ (red), i.e. from a less to a more quantum regime of the intracavity field. Each curve smoothly interpolates between the lower and the upper branch of the bistability loop with a crossover region located around the center of the bistability interval. The closer to the classical limit $g/\kappa\ll 1$, the sharper the crossover.

This behavior can be physically understood by looking at the Wigner function of the cavity field~\cite{vogel1988quantum} or, equivalently, at the photon number probability distribution in the steady state.  
The former is the Fourier transform
\begin{equation}
W_{ss}(\alpha) = \int d^2\eta\, D(\eta) e^{\alpha\eta^*-\alpha^*\eta}
\end{equation}
of the characteristic function $D(\eta)$ defined as 
\begin{equation}
 D(\eta) = \text{Tr}[\rho_{c,ss}\hat{d}(\eta)]   
\end{equation}
in terms of the displacement operator $\hat{d}(\eta)=\exp(\eta a^\dagger-\eta^*a)$ of displacement parameter $\eta$~\cite{walls2008quantum}. The latter is instead obtained by projecting the steady-state density matrix $\rho_{c,ss}$ onto a Fock state of number $n$,
\begin{equation}
P(n)={\rm Tr} [\rho_{c,ss} |n\rangle \langle n|]\,.
\end{equation}
Plots of these quantities are shown in Fig.\ref{fig:fig2}.(c) and Fig.\ref{fig:fig2}.(d) for a coherent drive amplitude $F_d$ tuned at the sharp jump of the quantum curve in Fig.\ref{fig:fig2}.(a).

Both quantities are characterized by a bimodal character, with a population distributed within a peak of high average photon number, identified as $m_{H}$, and within another one, of low photon number, identified as $m_{L}$. The positions of these two peaks roughly correspond to the steady-states of the mean-field approximation discussed above and both the photon number and the Wigner distributions are strongly suppressed in between the peaks. When the coherent drive amplitude $F_d$ is varied, the position of the peaks slowly moves following the mean-field steady states, but more importantly the relative weight of the two peaks dramatically changes. 

For a generic state, we can use the average photon number value as a threshold $\bar n\equiv (n_{H,MF}+n_{L,MF})/2$ to separate the cavity field state into the two components $m_H$ and $m_L$ as shown in Fig.\ref{fig:fig2}.(c,d), and define the probabilities  $P_{H(L)}$ to find the system in the $m_{H(L)}$ state in terms of the photon number distribution $p(n,t)={\rm Tr} [\rho(t) |n\rangle \langle n|]$ as $P_{L(H)}(t)=\sum_{n\lessgtr \bar n}p(n,t)$. As the photon number distribution is strongly suppressed in between the two peaks, the values of $P_{L(H)}$ are weakly dependent on the specific choice of the separating point $\bar n$.

The probabilities $P_{L,H}$ have comparable values in the crossover region while the weight gets quickly concentrated in the upper (lower) peak $P_{H(L)}$ for growing (decreasing) coherent drive amplitudes $F_d$. Physically, this result can be interpreted as the quantum dynamics allowing jumps between the two mean-field steady-states, but tending to favoring one of the two states depending on the coherent drive amplitude $F_d$.

The link to bistability is further clarified by looking at the Liouvillian gap $\Lambda$ defined as the lowest non-zero eigenvalue of the Liouvillian operator \eqref{eq:master_eq} and plotted as a dashed line in Fig.\ref{fig:fig2}(b) (dashed lines). The Liouvillian gap $\Lambda$ is a purely real quantity that describes the slowest timescale of the system relaxation towards the steady-state. While the finite value of $\Lambda$ confirms that the mean-field steady-states are not true steady-states in the bistable regime, its small value means that the relaxation timescale, $\Lambda^{-1}$, can be far slower than the microscopic relaxation dynamics $\kappa^{-1}$ in the central region of the bistability window $[F_-,F_+]$, as experimentally investigated in~\cite{Rodriguez_2017,Fink2018}. In the following, we will note $F_m$ the drive amplitude at which $\Lambda$ is minimum.

The slowing down effect and the depth of the minimum in $\Lambda$ gets more and more dramatic as the interaction strength $g$ is reduced and a larger number of photons are needed to observe nonlinear effects~\cite{Drummond_1980,vogel1988quantum}. In this classical limit, the strength of quantum fluctuations is suppressed and the dynamics bears close analogies to metastable effects at a first order phase transition \cite{Ciuti_1pdrive}: if the system is observed on a timescale shorter than the inverse Liouvillian gap $\Lambda^{-1}$, the full quantum dynamics recovers an effectively bistable behavior between the two mean-field steady-states. Relaxation to the true quantum steady-state only occurs on the longer time scale $\Lambda^{-1}$.

In preparation to the next Sections, it is useful to connect the Liouvillian gap $\Lambda$ to the spontaneous jump-up transition rate $\Gamma_{LH}$ from the state $m_{L}$ to $m_{H}$ and the spontaneous jump-down rate $\Gamma_{HL}$ from the state $m_{H}$ to $m_{L}$. 
 The jump dynamics is then summarized by the kinetic equation for $P_{L}(t)$
\begin{equation}
\frac{dP_{L}}{dt}=\Gamma_{HL}P_{H} - \Gamma_{LH}P_L\, 
\end{equation}
and the analogous one for $P_H(t)=1-P_L(t)$. With simple manipulations of these kinetic equations, the spontaneous jump-up and jump-down rates can be related to the steady-state probabilities $P_{L,H}^{ss}$ and the Liouvillian gap via
\begin{equation}
\Gamma_{LH (HL)} = P^{ss}_{H(L)}\,\Lambda\,.
\end{equation}
The thus calculated rates $\Gamma_{LH}$ (solid lines) and $\Gamma_{HL}$ (dotted lines) are shown in Fig.\ref{fig:fig2}(b) as a function of the coherent drive amplitude $F_d$. Correspondingly to the fact that the population is concentrated in the low $L$ (or high $H$) state, the jump down rate $\Gamma_{HL}$ (or the jump up rate $\Gamma_{LH}$) dominates in the left side $F_d<F_m$ (or in the right side $F_d>F_m$) of the Liouvillian gap minimum. On both sides, the dominating rate is approximately equal to the Liouvillian gap, while the other one is way smaller. For $F_d\sim F_m$ around the minimum, the populations are almost equal and the Liouvillian gap interpolates between the two spontaneous jump rates. 

\section{Preparation of the initial state and spontaneous jump-up dynamics}
\label{sec:prep-and-jump}

As a preliminary step in our study of the response to a single photon, we need to first establish a protocol capable of preparing the system in the desired initial state and characterize the system evolution in the absence of the incident single photon. As a most relevant configuration, we focus on the case where the system is initially prepared in the $m_L$ state as accurately as possible, so that the arrival of single incident photon may trigger a jump to the higher $m_H$ state.

As we have seen in the previous Section, the initial $m_L$ state is in general not a steady-state of the system. As such, it needs to be prepared over a time interval shorter than the spontaneous jump-up rate $\Gamma_{LH}$. At the same time, the preparation has to be slow enough to prevent the non-adiabatic creation of undesired excitations around $m_L$ during the preparation stage: this requires the ramp-up time to be longer than the inverse frequency of the collective Bogoliubov excitation on top of the $m_L$ state~\cite{Carusotto_2013,Claude:PRL2022}. For the values of detuning under consideration, the Bogoliubov frequency is on the order of the decay time $\kappa^{-1}$ or shorter. In order to ensure an accurate preparation of the $m_L$ state, a further clean-up stage of duration equal to a few $\kappa^{-1}$ is introduced after the end of the ramp, during which any leftover excitation gets exponentially damped at a rate $2\kappa$. This preparation protocol is summarized in Fig.\ref{fig:fig3}.(h).

The different panels of Fig.\ref{fig:fig3}.(a-g) illustrate the different aspects of the dynamics of the system after this preparation, in the absence of incident photon. Following the discussion in the previous Section, two very different regimes can be identified for the rate $\Gamma_{LH}$ of spontaneous jumps from the lower $m_L$ to the higher $m_H$ state. 

For $F_d<F_m$ far on the left side of the bistable interval -- referred to as the $LT$-regime -- the prepared initial state is very close to the steady-state state and the spontaneous jump-up rate $\Gamma_{LH}$ is practically negligible. 
This lack of spontaneous dynamics is illustrated in Fig.\ref{fig:fig3}(d-f) where we show snapshots of the time-evolution of the Wigner function $W_0(\alpha,t)$ for three relevant times, namely, right after preparation (d), a moderate time later (e), and a long time later (f). No appreciable time-evolution is visible in any panels and the Wigner function retains the characteristic circularly-symmetric shape of a low intensity coherent state with almost standard quantum fluctuations and no appreciable squeezing. The absence of population beyond the dashed circle shows that no spontaneous jumps to the $m_H$ state occur either and the temporally constant shape implies that that no dynamics due to collective excitations around the $m_L$ state is occurring either. The overall absence of evolution after preparation confirms that in the $LT$ regime our preparation protocol provides an accurate realization of the desired initial state.

For $F_d>F_m$ on the right side of the bistable interval -- referred to as the $RT$-regime -- the steady-state is dominantly $m_H$-like, so that the prepared $m_L$ state is only metastable with a sizeable rate $\Gamma_{LH}$ for spontaneous jump-up processes towards the $m_H$ state even in absence of the incident single photon. Snapshots of the Wigner function $W_0(\alpha,t)$ at different times during this spontaneous dynamics are shown in Fig.\ref{fig:fig3}(a-c).
The first snapshot is taken right after the preparation and shows that the system is predominantly $m_L$-like as intended. While the spontaneous jump-up process is already taking place, the weight of the $m_H$ peak is still $>100$ times smaller than that of $m_L$, so that this protocol can be reliably adopted for an accurate preparation of the initial state to be used in the next Sections.

In the following snapshot (b), the field is halfway throughout the spontaneous jump-up dynamics, as is shown by the balanced bimodal character of the Wigner function.
In the later-time snapshot (c), the system has almost reached its steady-state, largely dominated by the $m_H$ peak. This spontaneous dynamics in the $RT$ regime is further illustrated in Fig.\ref{fig:fig3}.(g). The yellow curve shows the evolution of the average intracavity photon number $\langle n(t)\rangle$ in time: as expected, this quantity grows and exponentially tends to its steady state $n_H$ on a time-scale set by the spontaneous jump-up rate $\Gamma_{LH}$. 

In contrast to this smooth evolution of the average photon number, the notion of a sudden jump-up process is clearly visible when considering individual quantum trajectories of the system within a Monte-Carlo wavefunction reformulation~\cite{breuer2002theory} of the master equation (\ref{eq:master_eq}). The green curves show a few examples of individual trajectories: each trajectory exhibits a sudden jump from $n_L$ to $n_H$ on a time-scale $\kappa^{-1}$. The specific instant at which each trajectory jumps varies stochastically from one trajectory to another: the spontaneous jump-up characteristic time $\Gamma_{LH}^{-1}$ observed in the average photon number curve corresponds to the spread of the instants at which each individual trajectory jumps.

\begin{figure*}[hbt]
\includegraphics[width=\textwidth]{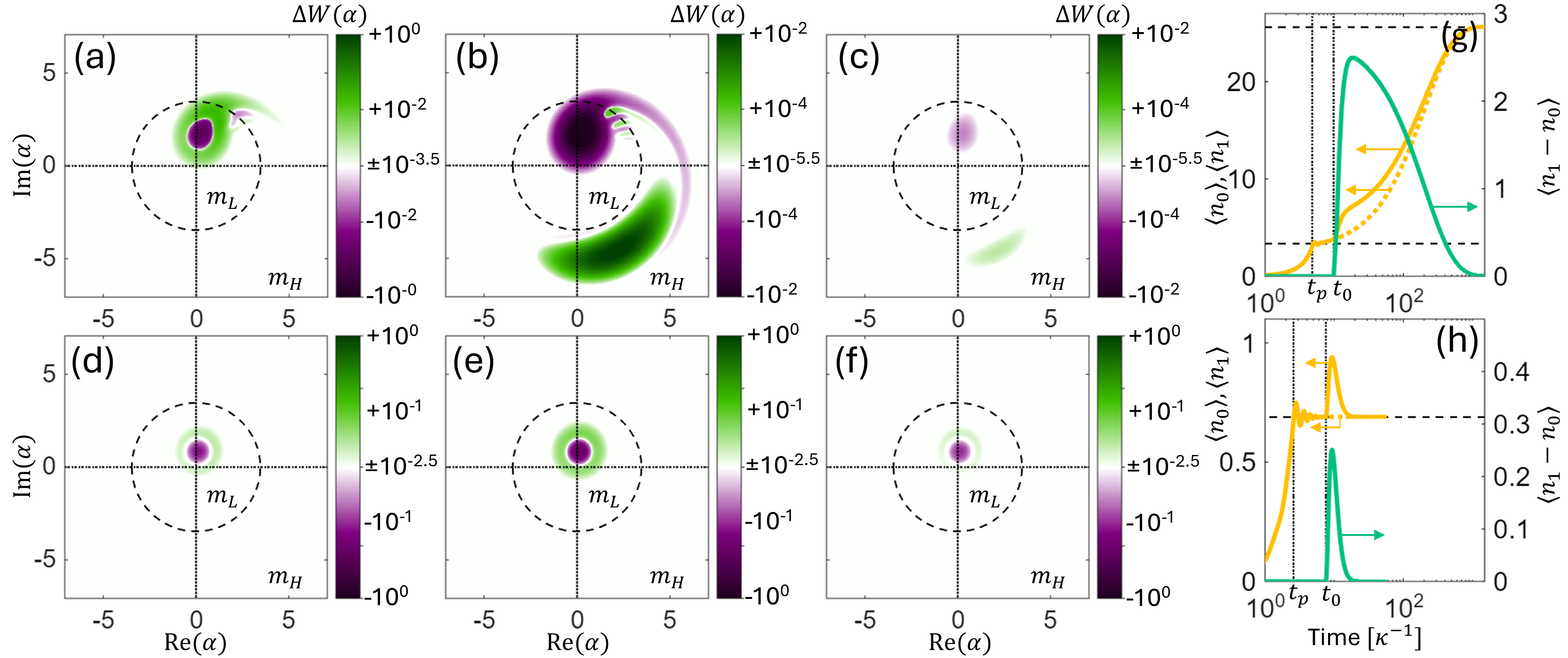}
\caption{Panels (a-f): Color-plots of the difference of the Wigner functions $\Delta W(t,\alpha)=W_{1}(t,\alpha)-W_{0}(t,\alpha)$ with and without the incident single photon. Panels (a-c) are in the $RT$-regime ($F_d/F_+=0.856$).
The snapshots are taken at times $t=t_0+1\kappa^{-1}=0.005\Lambda^{-1}$ (a), $t=t_0+100\kappa^{-1}=0.5\Lambda^{-1}$ (b), and $t_c=1500\kappa^{-1}=75\Lambda^{-1}$ (c). Panels (d-f) are in the $LT$-regime ($F_d/F_+=0.461$). The snapshots are taken at times $t=t_0+0.5\kappa^{-1}=0.15\kappa_e^{-1}$ (d), $t_c=t_0+2\kappa^{-1}=0.3\kappa_e^{-1}$ (e), and $t_c=5\kappa^{-1}=1.5\kappa_e^{-1}$ (f). Panels (g-h): plots of the expectation value of the intracavity photon number in the presence $n_1(t)$ (solid yellow line) and in the absence $n_0(t)$ (dashed yellow line) of the incident single photon. The difference $n_1(t)-n_0(t)$ is shown as the green solid line (right y-axis). Panel (g) is for $F_d/F_+=0.856$ in the $RT$ regime and panel (h) is for $F_d/F_+=0.461$ in the $LT$ regime. Same parameters as in Fig.\ref{fig:fig3} and $\kappa_e=0.3\kappa$.}
\label{fig:fig4}
\end{figure*}

\section{Response to an incident single photon}
\label{sec:jump-with-pert}

We now proceed to investigate how this spontaneous jump-up dynamics is modified when a single photon hits the left mirror of the cavity at $t=t_0$. This is done using the cascaded quantum systems method discussed in Sec.\ref{sec:system}: the incident single-photon wavepacket is created via spontaneous emission by a two-level system: its temporal length is set by the decay rate $\kappa_e$ of the two-level system, while its frequency $\omega_e$ is tuned on resonance with the Bogoliubov excitations mode around the $m_L$ metastable state (see App.~\ref{app:ph_freq}).

\subsection{Single-photon-stimulated jump-up dynamics}

Let us first focus on the $RT$-regime $F_d>F_m$. In Fig.~\ref{fig:fig4}.(a-c) we plot the difference $\Delta W(\alpha)\equiv W_1(\alpha)-W_0(\alpha)$ between the Wigner functions obtained with ($W_1(\alpha)$), and without ($W_0(\alpha)$) the incident single photon. Right after the arrival of the photon [panel (a)],  
the Wigner function difference $\Delta W$ shows a broadening of the $m_L$ peak due to the excitation around the $m_L$ metastable state and a simultaneous triggering of the jump-up process as indicated by the comma-like pattern whose appearance is anticipated by the incident single-photon. The jump-up processes induced by the incident photon show up even more clearly at a slightly later time [panel (b)] 
where the reinforced $m_H$ peak is the signature of the incident-photon-stimulated  jump-up processes. At late times, the effect of the single-photon disappears as the system approaches the same steady-state as the one without incident photon.

This dynamics is further illustrated in Fig.\ref{fig:fig4}(g) where the solid (dashed) green lines show the temporal evolution of the average photon number $n_1(t)$ ($n_0(t)$) in the presence (absence) of incident photon. While the $n_0(t)$ curve shows the smooth approach to the steady-state on a time-scale set by $\Gamma_{LH}$ discussed above, the curve for $n_1(t)$ features an additional fast increase around the arrival time of the photon. This fast increase is even clearer on the green line showing the difference $n_1(t)-n_0(t)$: the maximum in the difference is reached within a short time-scale {set by the transit time $\tau_{tr}$ (defined in App.~\ref{app:res_T} and satisfying $\min[\kappa_e^{-1},\kappa^{-1}]<\tau_{tr}<\max[\kappa_e^{-1},\kappa^{-1}]$) that the photon spends inside the cavity and hence perturb the intracavity field dynamics}. After the transit time $\tau_{tr}$, the difference $n_1(t)-n_0(t)$ then persists for a long time $\Gamma_{LH}^{-1}$ set by the spontaneous jump-up processes. 

As the in-cavity photon number is proportional to the transmitted light intensity, this plot highlights how a single incident photon is able to trigger jump-up processes that can be then optically read out as a long-lasting increase of the transmitted light. A quantitative study of the photo-detection efficiency and a comparison to dark-counts due to spontaneous jump-up processes will be given in the next Section.

\subsection{Ultrafast single-photon-enhanced transmission}

A quite different behavior is predicted in the $LT$ regime for $F_d<F_m$. Here, the steady state is dominated by the $m_L$ component and the system mostly returns to it after the perturbation has gone. This behavior is illustrated in the snapshots of the evolution of the Wigner function difference in response to an incident single-photon shown in Fig.~\ref{fig:fig4}.(d-f): the incident single photon is able to significantly perturb the state of the system, broadening the Wigner distribution around the $m_L$ state. This  remains however localized within the dashed circle and no significant single-photon-stimulated jump-up processes into the $m_H$ state occur.

This dynamics is summarized in the plots  of the photon number $n_{1}$ ($n_0$) in the presence (absence) of incident photon and of their difference shown in Fig.~\ref{fig:fig4}.(h): even in the absence of a single-photon-stimulated jump-up process, the incident photon is able to trigger a sizable increase in the photon number that can be detected as an ultrafast pulse of enhanced transmission of the coherent drive beam. The duration of this pulse is much faster than the slow jump dynamics set by $\Gamma_{LH,HL}$ and it is instead determined by the transit time $\tau_{tr}$ of the incident photon introduced above and defined in App.~\ref{app:res_T}.

\begin{figure}[hbt]
\includegraphics[width=\columnwidth]{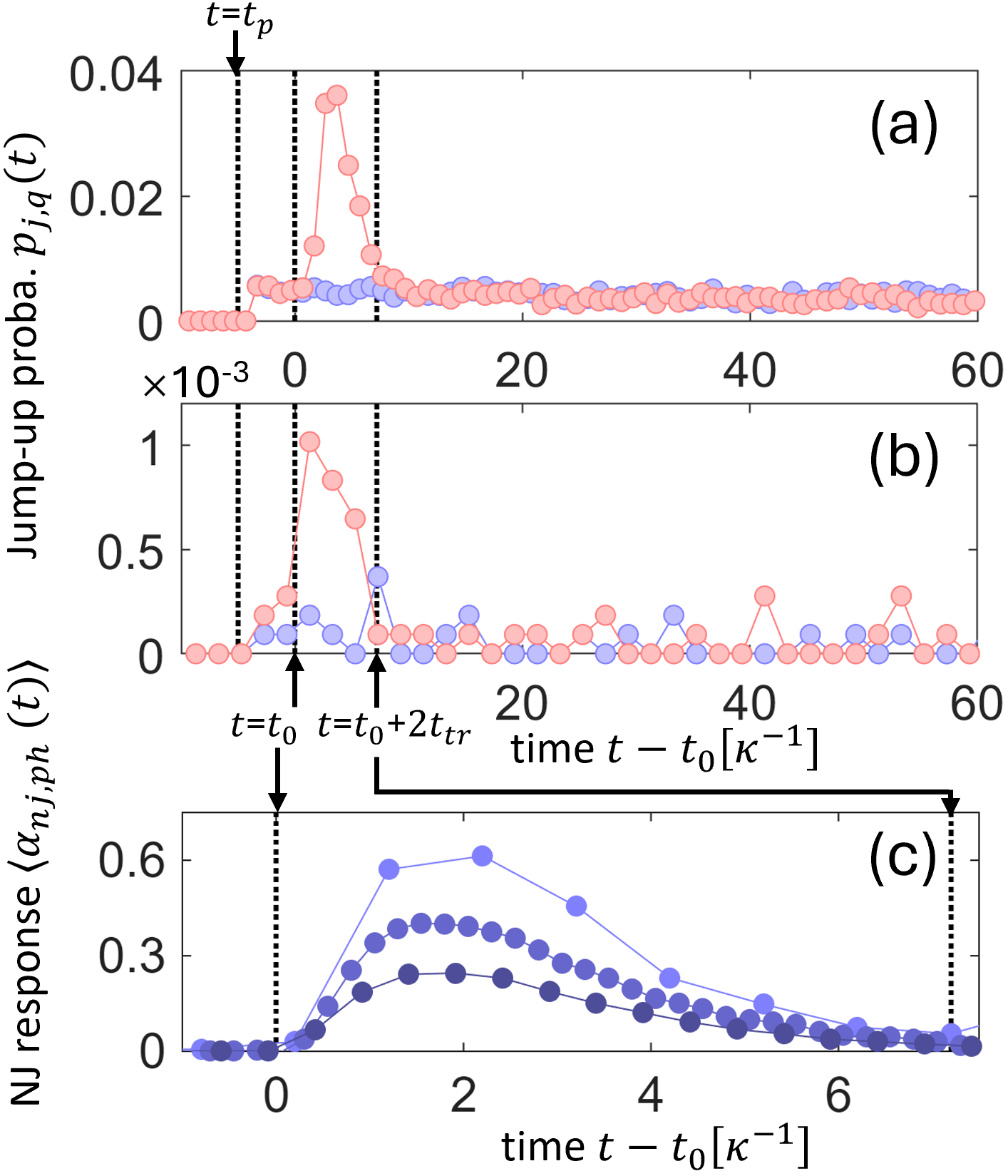}
\caption{Panels (a,b): plots of the jump-up probability $p_{j,q}(t)$ within a time interval $t\pm0.5\kappa^{-1}$ in the presence (red symbols, $q=1$) and in the absence (blue symbols, $q=0$) of the single photon perturbation for $F_d/F_+=0.856$ (a) and $0.750$ (b) in respectively the $RT$ and the $LT$ regimes.
Panel (c): plot of the average number $\alpha_{nj,ph}(t)$ of additional intra-cavity photons in response to the incident single photon within the subset of events in which no jump occurred within $[t_0,t_0+2t_{tr}]$. Total number of realizations $\mathcal{N}_r=10^4$. The three datasets are for $F_d/F_+=0.856$ (light blue symbols), $0.750$ (blue symbols) and $0.461$ (darker blue symbols). The time $t_0$ is the arrival time of the incident single photon. System parameters $g=0.2\kappa$,
$\Delta\omega=-8\kappa$ and $\kappa_e=0.3\kappa$. }
\label{fig:fig5}
\end{figure}

\subsection{Analysis of the response in terms of quantum trajectories}
\label{sec:trajectory}

These two different behaviors can be further characterized by looking at the individual quantum trajectories of a Monte Carlo wavefunction calculation~\cite{breuer2002theory}. We first focus on the jump-up response. In either the presence ($q=1$) or absence ($q=0$) of the incident photon, we consider the probability distribution $p_{j,q}(t)$ that the system jumps up within a short time interval $t\pm0.5\kappa^{-1}$ as a function of the time $t$. For each trajectory, we identify the jump time $t$ -- if any -- as the time at which the photon number crosses $\bar n$ and remains above it for the next time step.  

The jump-up probability $p_{j,q}(t)$ obtained in this way by averaging over a large number of realization is plotted in Fig.\ref{fig:fig5}(a,b) for two values of the  coherent drive amplitude $F_d$ within the $RT$ regime. Panel (a) is for a value of $F_d$ close to the right edge of the hysteresis loop, while panel (b) is for a pump strength close to the middle of it. 
In both cases, the probability $p_{j,1}(t)$ in the presence of the incident photon displays a strong peak on top of a flat background set by $p_{j,0}(t)$, i.e the spontaneous jump-up probability at time $t$ in the absence of the photon. As it is indicated by the middle and right vertical dashed lines, the peak starts right after the arrival time $t_0$ of the photon and lasts for a time approximately given by {the transit time $\tau_{tr}$}. In the $LT$ regime, instead, the $m_L$ state is very stable and no jump up events are found in the trajectories, consistently with the negligible spontaneous jump-up rate $\Gamma_{LH}$. This analysis confirms our intuitive picture that the incident single photon is able to stimulate the  jump-up process into the $m_H$ state, and that this stimulation occurs within the photon transit time $\tau_r$.

Besides triggering jump up events, the incident photon can have an effect also on the dynamics of the system around the $m_L$ state. In the previous subsection, we have seen such behavior for the $LT$ regime where jumps do not take place. Now, we make use of the trajectory picture to show that such a behavior is also present in the $RT$ regime but is typically hidden by the jump-up processes. To highlight it, we consider the average increase of photon number $\langle \alpha_{nj}(t)\rangle = \langle n_{nj,1}(t)-n_{nj,0}(t) \rangle$ upon restricting the average to those trajectories that do not display any jump within the interval $T_p$. This allows to remove from the curves of Fig.\ref{fig:fig4}.(g) the dominant contribution of the jump up process. 

The result of such calculations is shown in Fig.\ref{fig:fig5}.(c) for three values of the pump strength in either the $LT$ (dark blue) and $RT$ regime (blue and light blue): in all cases, it is apparent how also the no-jump trajectories display a marked increase of the photon number as a consequence of the incident single photon and that this increase lasts for a time approximately given by the transit time $\tau_{tr}$. From a quantitative point of view, the increase appears to be larger in the $RT$ regime but the variation across the bistability region is however not dramatic.

\begin{figure}[h]
    \centering
    \includegraphics[width=\columnwidth]{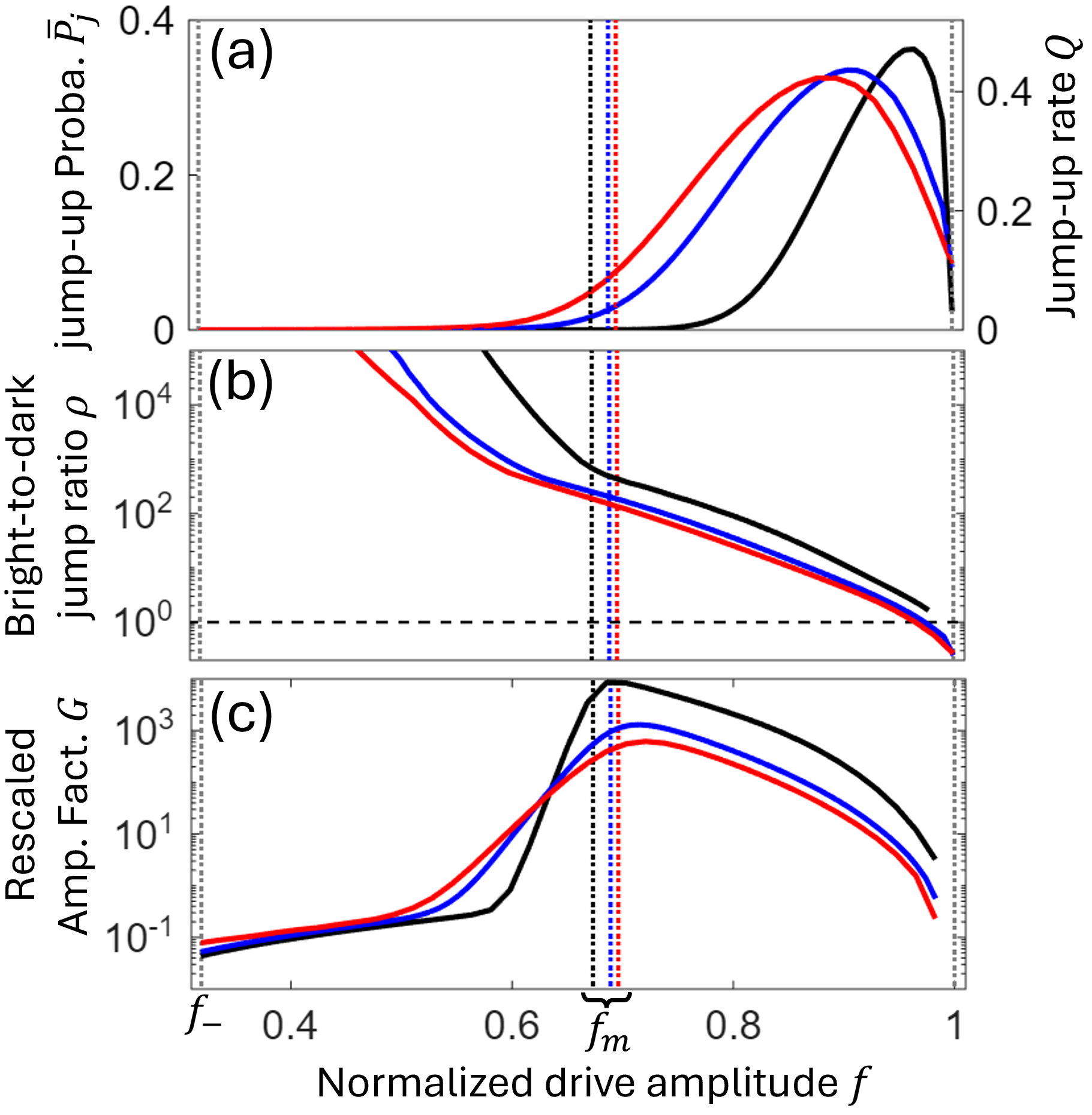}
    \caption{Figures-of-merits of the response to the incident single photon as a function of the normalized drive amplitude $f=F_d/F_+$. Panel (a) shows the single-photon-stimulated jump-up probability $\bar P_j$ (left axis), and the single-photon-stimulated jump-up rate $Q$ (right axis). Panel (b) shows the ratio of single-photon-stimulated (bright) to spontaneous (dark) jump-up rates $\rho=Q/\Gamma_{LH}$. Panel (c) shows the rescaled amplification factor $G$ defined in \eqref{eq:defG} as a sort of photo-multiplication factor. In all panels, the black, blue and red curves correspond to different strengths of the nonlinearity $g/\kappa=0.2,\,0.4,\, 0.5$. The left (right) gray dashed line shows the left (right) turning points $F_-$ ($F_+$), while the colored dashed lines indicate the position $f_m=F_m/F_+$ of the Liouvillian gap minimum for the three considered values of $g/\kappa$. Calculation parameters: $\Delta\omega=-8\kappa$, $\kappa_e=0.3\kappa$.}
    \label{fig:fig6}
\end{figure}

\section{Quantitative analysis of the cavity response to a single photon}
\label{sec:analysis}

In the previous Section we have given a qualitative discussion of the different effects that take place in the cavity in response to the incident single photon. Here we proceed with a quantitative characterization of the magnitude of these effects using appropriately designed figures of merit. 

\subsection{Jump-up probability}

As a first natural figure of merit, we consider the probability $\bar P_j$ of single-photon-stimulated jump-up processes. This can be extracted (see Appendix \ref{app:Pj} for a full derivation) from 
the temporal evolution of the probability $P_{H,L}(t)$ to find the cavity field in the $m_{H,L}$ states according to the definition in Sec.\ref{sec:quantumth}. The jump probability can be obtained by evaluating 
\begin{equation}
\bar P_j=\max_t\left[P_j(t)\right]=
   \max_t \left[\frac{P_H^{(1)}(t)-P_H^{(0)}(t)}{1-P_H^{(0)}(t)}\right]\,.
    \label{eq:P_phot_jump}
\end{equation}
where the index $q$ in $P^{(q)}_{H}$ respectively refers to the case with ($q=1$), or without ($q=0$) the incident photon. In the photodetection vocabulary, this jump-up probability can be understood as a photo-detection efficiency. In the quantum trajectory picture of Sec.\ref{sec:trajectory}, the quantity $P_j(t)$ is related to the detection probability rate $[p_{j,1}(t)-p_{j,0}(t)]\kappa$, as $\partial P_j/\partial t \simeq [p_{j,1}(t)-p_{j,0}(t)]\kappa$.

A plot of the jump-up probability $\bar P_j$ as a function of $F_d$ across the entire bistability window from the $LT$ to the $RT$ regime is shown in Fig.\ref{fig:fig6}.(a) (right axis) for three different values of the nonlinearity strength $g/\kappa=0.2$ (black), $0.4$ (blue) and $0.5$ (red). For all curves, the frequency of the incident photon is taken at the peak of the response resonance, which roughly coincides with the frequency of the Bogoliubov mode around the $m_L$ state (see the black curve in Fig.\ref{fig:fig8} of Appendix \ref{app:ph_freq}).

Fig.\ref{fig:fig6}.(a) shows how the single-photon-stimulated jump-up response to the single photon increases sharply with increasing $F_d$ starting from the left edge of the bistability loop. The jump-up probability also strongly grows when the nonlinearity parameter $g/\kappa$ is increased towards the quantum regime (black to blue to red curves): indeed, the smaller the number of photons involved in the dynamics, the stronger the response to an additional single photon.

However, all curves reach a maximum value well before the right edge of the bistability window. For larger $F_d$ past this maximum, the single-photon-stimulated jump-up probability goes back down. This is due to the spontaneous jump-up processes that deplete the initial $m_L$ state faster than the characteristic photon transit time $\tau_{tr}$. This interpretation is validated by the sharp growth of the spontaneous jump-up rates $\Gamma_{LH}$ as the edge of the bistability loop is approached, as shown in Fig.\ref{fig:fig2}.(b). As expected, this competing effect is more important and the maximum of $\bar P_J$ is reached for lower $F_d$ when the nonlinearity parameter $g/\kappa$ is increased towards the quantum regime. The overall maximum value of the jump-up probability is then determined by a trade-off of these different effects.

\begin{figure}
    \centering
    \includegraphics[width=\columnwidth]{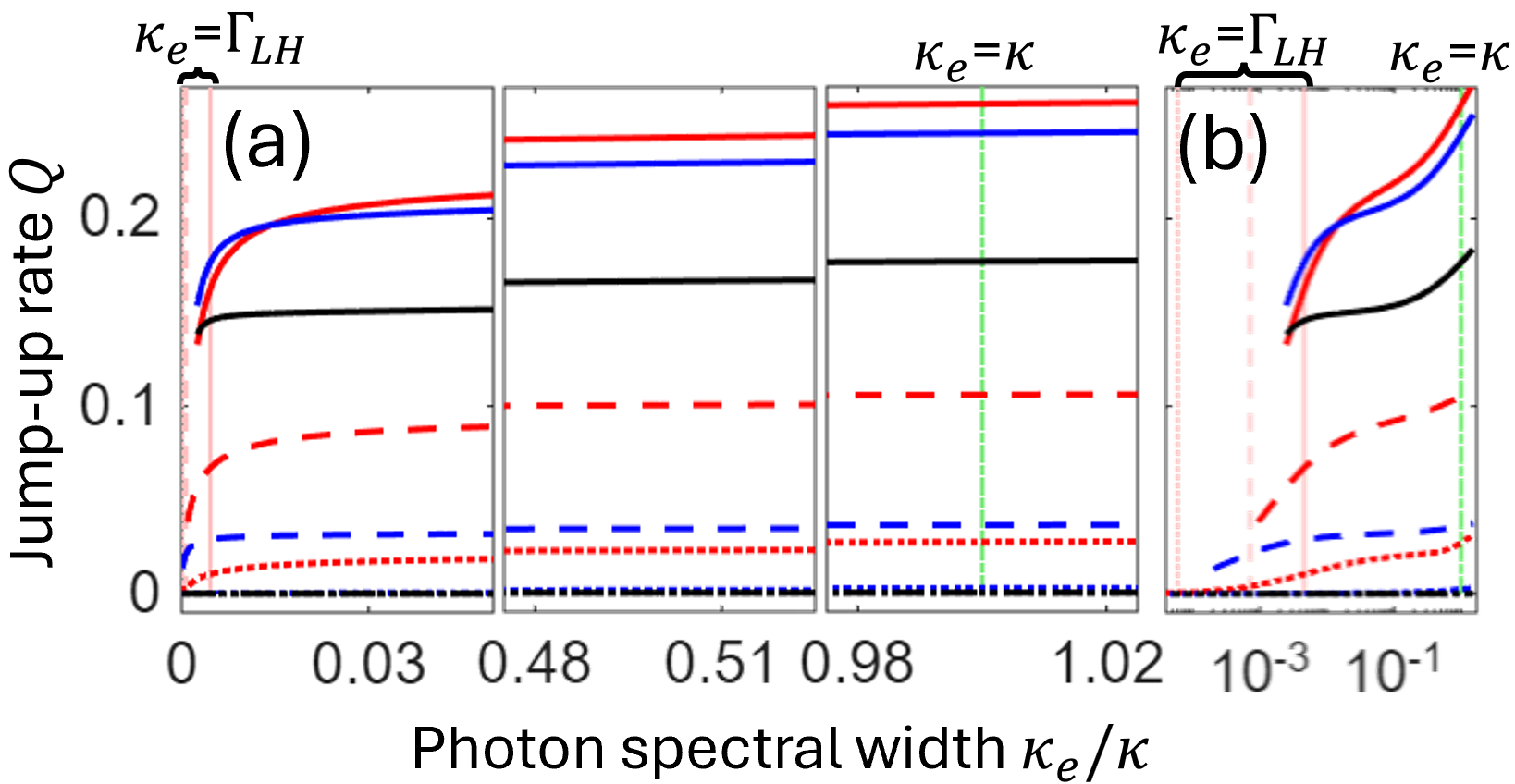}
    \caption{Panel (a): plots of the single-photon-stimulated jump-up rate $Q$ as a function of the incident photon linewidth $\kappa_e$ at resonance. the red, blue and black lines refer to $g/\kappa=0.5$, $g/\kappa=0.4$, $g/\kappa=0.2$, respectively. The three line styles refer to different values of the coherent drive amplitude regime:  dotted line for the $LT$-regime ($f=0.627$,$f=0.572$, $f=0.462$, for $g/\kappa=0.5$, $g/\kappa=0.4$, $g/\kappa=0.2$ respectively), dashed line for $f=f_m$ ($f=0.696$,$f=0.689$, $f=0.673$), and solid line in the $RT$-regime ($f=0.761$,$f=0.789$, $f=0.856$). The three panels show three different intervals: around $\kappa_e=\Gamma_{LH}$ (left), around $\kappa_e=0.5\kappa$ (center), and around $\kappa_e=\kappa$ (right). The pink vertical lines in the left panel and in panel (b) show $\kappa_e=\Gamma_{LH}$ for $g/\kappa=0.5$ for all three values of $f$, with the line styles as defined before. The green dash-dotted vertical line in (a)'s central panel and in panel (b) show the case for $\kappa_e=\kappa$. Panel (b) shows the full datasets on an horizontal log scale. Calculation performed for $\Delta\omega=-8\kappa$.}
    \label{fig:fig7}
\end{figure}

\subsection{Jump-up rate}

To disentangle the different effects contributing to $\bar{P}_j$, it is interesting to write this quantity as the product $\bar{P}_j=QT_{res}$ of two contributions, the residence time $T_{res}$ of the photon in the cavity and a single-photon-stimulated jump-up rate $Q$. The photon residence time is defined to capture both the effectiveness of the injection of the incident photon into the cavity via the left mirror and the duration of the time interval it actually spends in the cavity. As the initial $m_L$ state is weakly affected by the nonlinearity, the residence time $T_{res}$ of a resonant incident photon can be estimated using the linear cavity model of App.~\ref{app:res_T} to be
\begin{equation}
T_{res}=\frac{2 \kappa_1}{\kappa(\kappa+\kappa_e)}\,.
\label{eq:T_res_fin}
\end{equation}
For a narrow photon spectrum $\kappa_e\ll \kappa$ (i.e. a temporally long photon wavepacket), the spectral overlap of the incident photon with the cavity mode is good and the residence time is approximately constant $T_{res}\simeq 2\kappa_1/\kappa^2$ and determined by the left mirror transmittivity and the cavity decay time $\kappa^{-1}$. For a broad photon spectrum $\kappa_e\gg \kappa$ (i.e. a temporally short photon wavepacket), the residence time $T_{res}\simeq {2\kappa_1}/{\kappa\kappa_e}$ is suppressed by the reduced spectral overlap.

Remarkably, this quite straightforward definition of $T_{res}$ allows to capture much of the $\kappa_e$-dependence of $\bar{P}_j$: as it is shown in Fig.\ref{fig:fig7}(a) for different values of $g/\kappa$, the single-photon-stimulated jump-up rate $Q$ shows in fact a quite weak dependence over a large interval of values of $\kappa_e$. This means that $Q$ can be understood as an internal single-photon-stimulated jump-up rate which characterizes the intracavity field response and weakly depends on the details of the incident photon wavepacket. A significant reduction of the jump-up response $Q$ only appears for very small $\kappa_e \lesssim \Gamma_{LH}$, due again to the 
spontaneous jump-up rate becoming faster than the photon intracavity transit time $t_{tr}$.

The fact that $Q$ is approximately constant for $\kappa_e\gg \Gamma_{LH}$ shows the importance of performing a full quantum theory. In a  semiclassical approach, a reduction of $\kappa_e$ far below $\kappa$ would lead to a vanishing intracavity intensity~\cite{Petrovnin:PRXQ2024}, and hence to an non-physical drop of the single-photon-stimulated jump-up probability. Our quantum theory is instead able to fully take this into account: owing to the discrete nature of the photon, decreasing $\kappa_e$ extends the time interval during which the jump-up may occur, but not its overall probability to occur $\bar{P}_j$.

In order to get an idea of the actual value of the single-photon-stimulated jump-up rate $Q$, the right $y$-axis of Fig.\ref{fig:fig6}.(a) indicates the value of this quantity across the bistability loop in units of the cavity decay rate $\kappa$. But an even more interesting comparison involves the competing spontaneous jump-up rate $\Gamma_{LH}$: the ratio $\rho=Q/\Gamma_{LH}$ quantifies in fact the actual observability of single-photon-stimulated jump-up processes on top of the background of spontaneous processes that play the role of dark counts in a detector language. 

This quantity is plotted in Fig.\ref{fig:fig6}.(b) for three different values of $g/\kappa$. Quite interestingly, a larger $\rho$ is achieved for smaller values of $g/\kappa$, i.e. approaching the mean-field limit: the probability of single-photon-stimulated jump-up processes drops slower than the one for spontaneous processes. Still, one needs to remember that, as shown in panel (a),  this comes at the price of a markedly lower single-photon-stimulated jump-up rate $Q$, i.e. a lower detection efficiency.

\subsection{Photo-multiplication factor}

Another important quantity characterizing the response of the cavity to the incident single-photon perturbation is its brightness, namely the number of photons added to the transmitted beam as a result of the incident single photon. This can be physically understood as the magnitude of the photon avalanche triggered by the photon.

This avalanche can be characterized by an amplification factor defined as
\begin{equation}
   \alpha=2\kappa_2 \int_0^{\infty} dt \,  [n_1(t)-n_0(t)]\,.
   \label{eq:alpha}
\end{equation}
As it includes the perturbing photon itself, this quantity is typically lower-bounded by $\alpha_{min}=2\kappa_2 T_{res}$. We thus define the rescaled amplification factor 
\begin{equation}
G=\frac{\alpha}{2\kappa_2 T_{res}}-1
\label{eq:defG}
\end{equation}
which is shown in  Fig.\ref{fig:fig6}.(c). 
By construction, this quantity is lower-bounded by zero, which corresponds to having no additional photon on top of the perturbing photon itself. Different effects contribute to $G$ in the different regimes. 

In the $RT$ regime, we have seen in Sec.\ref{sec:trajectory} that the additional photons due to no-jump processes are at most of the order one, while the massive contribution to $G$ that is visible in Fig.\ref{fig:fig7}.(c) comes from the single-photon-stimulated jump-up processes. The average number of additional transmitted photons given by a single-photon-stimulated jump-up process can in fact be estimated to $\kappa_2(n_{H,MF}-n_{L,MF})/\Gamma_{LH}$: as the effect of a jump-up lasts for a macroscopic time $\Gamma_{LH}^{-1}$, the number of extra photons grows when $g/\kappa$ is decreased towards the mean-field limit (red to blue to black curve), overcompensating the corresponding decrease of the single-photon-stimulated jump probability $\bar{P}_j$. The same happens when $F_d$ is decreased towards the center of the bistability loop: there are less single-photon-stimulated jumps, but each adds much more photons. This result in $G$ achieving its maximum value around the center of the bistability loop, i.e. around $F_d=F_m$.

Deep into the $LT$ regime, the single-photon-stimulated jump-up probability is instead negligible and the photo-multiplication factor $G$ is dominated by the no-jump processes. As such, the photo-multiplication factor $G$ is quite moderate but, interestingly, it gets larger as one moves deeper in the quantum limit at large $g/\kappa$. This is likely the result of a nonlinear correction to the effective detuning of the coherent drive field, that scales like $\Delta\omega_{nl} \simeq\Delta\omega+g(\langle a^\dagger a\rangle-1)$). When the field is in the $m_L$ state, $\Delta\omega_{nl}\simeq \Delta\omega<0$ so that the addition of a single photon shifts $\Delta\omega_{nl}$ closer to resonance by an amount $g$, which results in an enhanced transmission during the photon transit time.

Most interesting, and worth future studies, is the region around $F_d\lesssim F_m$ that in recent works~\cite{Ciuti_1pdrive} was associated to the critical point of the phase transition. For $F_d<F_m$, the steady-state is dominated by the $m_L$ state, so at late times after the arrival of the photon the system eventually relaxes back to this state. For suitable values of $g/\kappa\sim 0.5$ (i.e. at the onset of the quantum regime), and a pump strength $F_d$ sufficiently close to $F_m$, an interesting compromise is found between having a significant probability of jumping up [Fig.\ref{fig:fig6}.(a)] and a slow rate $\Gamma_{HL}$ for the decay back to the initial state. Together, the sizable probability of undergoing a single-photon-stimulated jump-up and the long time spent in the $m_H$ state, 
contribute to give a very large value of the photo-multiplication factor $G$, with the
additional
advantage in the context of photo-detection, that the system automatically returns to the initial state $m_L$, ready for the next photodetection event.

\section{Conclusions and outlook}
\label{sec:conclusions}

In this work, we have developed a fully quantum theory of the response of a coherently-driven Kerr cavity to a single incident photon resonant with the main excitation mode of the system. The cavity is initially prepared into the low-intensity state $m_L$ of the optical bistability loop and the single photon is able to trigger jumps towards the high-intensity state $m_H$. Consistently with the physical picture of a system sitting at the transition edge between two macroscopically different states $m_{L,H}$, a strong response is found in suitable regimes, largely exceeding the strength of the original perturbation. Our fully quantum approach allows us to obtain a detailed physical picture of the response and identify the different processes at play depending on the operating point within the bistability window. 

Two regimes that appear most promising in view of photo-detection applications are specifically identified and their performance quantitatively analyzed. At high values of the coherent drive amplitude $F_d$ on the right side of the bistability window, the prepared state $m_L$ is only metastable and the single photon has a significant probability -- a sort of quantum efficiency -- of stimulating a jump-up event into $m_H$: up to $30\%$ at the optimal driving point in $F_d$. We then find that the resulting ratio $\rho$ of the stimulated vs. spontaneous jump-up rates --a sort of bright to dark count ratio in the photo-detection language-- can exceed a value of order 10 at this optimal point. Far larger values of $\rho$ can be attained for lower values of $F_d$, at the price of a reduced quantum efficiency.

When a jump-up process occurs, it is associated to a strong increase of the transmitted intensity of the coherent drive beam.  Thanks to the very slow jump-up and jump-down rates as compared to the intrinsic cavity decay rate $\kappa$, the amplification  factor $G$ -- a sort of photo-multiplication factor--  can reach values on the order of tens to thousands depending on the chosen value of $F_d$. 

When the system is driven on the right part of the bistability window, this response is irreversible and, after a jump-up event, the system remains in the $m_H$ state and is no longer able to respond to following photons. The response can be made reversible by tuning the coherent drive amplitude in the center-left part of the bistability window where the system is able to automatically recover the initial $m_L$. Overall, this regime comes with several other interesting features: a still significant rate for single-photon-stimulated jump-up processes, to be compared to a much weaker rate for spontaneous jump-up processes, as well as very large photo-multiplication factor resulting from the slowest jump rates. 
Throughout all regimes, it appears that moderate values of the nonlinearity $g/\kappa\sim 0.2 \div 0.5$ on the edge of the quantum regime are a most promising choice for all the figures-of-merit of our proposed single-photon avalanche photo-detector.

Our study was carried out having in mind experimental implementations based on semiconductor technology for visible or near-infrared light, in particular nonlinear micropillar cavities in which the bistable regime is well characterized ~\cite{PhysRevA.69.023809,Carusotto_2013} and state-of-the-art devices are approaching the required values of the nonlinearity~\cite{Togan_2018,Rosenberg_2018,Tan_2020,Munoz_2019,Delteil_2019,Datta_2022}. However, our strategy and results are fully general and can be straightforwardly translated to any other wavelength domain where the required nonlinearity is available. The development of single-photon photo-detectors in the microwave domain is for instance an active and challenging topic, of paramount importance for superconducting circuit quantum technologies~\cite{Inomata_2016,Petrovnin:PRXQ2024,Lescanne:2020,Besse:2018,Oppliger:2025,matern2025dispersive,zeller2026stroboscopic}: the implementation of our proposal in this spectral domain using Josephson junctions as the nonlinear element of quantum circuits~\cite{blais2021circuit} appears to be a very exciting perspective for future work.

From a theoretical point of view, 
a natural next step will be to investigate the response of spatially-extended devices, such as planar microcavities, driven by a spatially extended beam. In analogy with first-order transitions in conservative systems, such as atomic ferromagnets~\cite{Zenesini2024}, ferromagnetic spin chains~\cite{Lagnese:PRB2021,Johansen:arXiv2025,Maertens:arXiv2025}, and cosmological systems~\cite{Coleman1977,Devoto_2022}, we anticipate that the jump-up process will occur via a bubble mechanism where an incident single photon generates an initially spatially localized bubble that then quickly expands in the plane. With a proper design of the system, we expect that this configuration may yield even higher quantum efficiencies and photo-multiplication factors.

\section*{Acknowledgements}
We are grateful to Gianluca Rastelli, Stephanie Matern and Simone Felicetti for continuous discussions. IC acknowledges financial support from the Provincia Autonoma di Trento; the Q@TN Initiative; the National Quantum Science and Technology Institute through the PNRR MUR Project under Grant PE0000023-NQSTI, co-funded by the European Union -- NextGeneration EU. This work is part of HQI (www.hqi.fr) initiative and is supported by France 2030 under the French National Research Agency grant number ANR-22-PNCQ-0002.


\begin{thebibliography}{68}%
\makeatletter
\providecommand \@ifxundefined [1]{%
 \@ifx{#1\undefined}
}%
\providecommand \@ifnum [1]{%
 \ifnum #1\expandafter \@firstoftwo
 \else \expandafter \@secondoftwo
 \fi
}%
\providecommand \@ifx [1]{%
 \ifx #1\expandafter \@firstoftwo
 \else \expandafter \@secondoftwo
 \fi
}%
\providecommand \natexlab [1]{#1}%
\providecommand \enquote  [1]{``#1''}%
\providecommand \bibnamefont  [1]{#1}%
\providecommand \bibfnamefont [1]{#1}%
\providecommand \citenamefont [1]{#1}%
\providecommand \href@noop [0]{\@secondoftwo}%
\providecommand \href [0]{\begingroup \@sanitize@url \@href}%
\providecommand \@href[1]{\@@startlink{#1}\@@href}%
\providecommand \@@href[1]{\endgroup#1\@@endlink}%
\providecommand \@sanitize@url [0]{\catcode `\\12\catcode `\$12\catcode
  `\&12\catcode `\#12\catcode `\^12\catcode `\_12\catcode `\%12\relax}%
\providecommand \@@startlink[1]{}%
\providecommand \@@endlink[0]{}%
\providecommand \url  [0]{\begingroup\@sanitize@url \@url }%
\providecommand \@url [1]{\endgroup\@href {#1}{\urlprefix }}%
\providecommand \urlprefix  [0]{URL }%
\providecommand \Eprint [0]{\href }%
\providecommand \doibase [0]{https://doi.org/}%
\providecommand \selectlanguage [0]{\@gobble}%
\providecommand \bibinfo  [0]{\@secondoftwo}%
\providecommand \bibfield  [0]{\@secondoftwo}%
\providecommand \translation [1]{[#1]}%
\providecommand \BibitemOpen [0]{}%
\providecommand \bibitemStop [0]{}%
\providecommand \bibitemNoStop [0]{.\EOS\space}%
\providecommand \EOS [0]{\spacefactor3000\relax}%
\providecommand \BibitemShut  [1]{\csname bibitem#1\endcsname}%
\let\auto@bib@innerbib\@empty
\bibitem [{\citenamefont {Huang}(2008)}]{huang2008statistical}%
  \BibitemOpen
  \bibfield  {author} {\bibinfo {author} {\bibfnamefont {K.}~\bibnamefont
  {Huang}},\ }\href {https://books.google.it/books?id=ZHl8HLk-K3AC} {\emph
  {\bibinfo {title} {Statistical Mechanics, 2nd Ed}}}\ (\bibinfo  {publisher}
  {Wiley India Pvt. Limited},\ \bibinfo {year} {2008})\BibitemShut {NoStop}%
\bibitem [{\citenamefont {Wilson}(1911)}]{Wilson_1911}%
  \BibitemOpen
  \bibfield  {author} {\bibinfo {author} {\bibfnamefont {C.~T.~R.}\
  \bibnamefont {Wilson}},\ }\href@noop {} {\bibfield  {journal} {\bibinfo
  {journal} {Proc. R. Soc. Lond. A Math. Phys. Sci.}\ }\textbf {\bibinfo
  {volume} {85}},\ \bibinfo {pages} {285} (\bibinfo {year} {1911})}\BibitemShut
  {NoStop}%
\bibitem [{\citenamefont {Glaser}\ and\ \citenamefont
  {Rahm}(1955)}]{Glaser_1955}%
  \BibitemOpen
  \bibfield  {author} {\bibinfo {author} {\bibfnamefont {D.~A.}\ \bibnamefont
  {Glaser}}\ and\ \bibinfo {author} {\bibfnamefont {D.~C.}\ \bibnamefont
  {Rahm}},\ }\href {https://doi.org/10.1103/PhysRev.97.474} {\bibfield
  {journal} {\bibinfo  {journal} {Phys. Rev.}\ }\textbf {\bibinfo {volume}
  {97}},\ \bibinfo {pages} {474} (\bibinfo {year} {1955})}\BibitemShut
  {NoStop}%
\bibitem [{\citenamefont {Irwin}(1995)}]{Irwin_1995}%
  \BibitemOpen
  \bibfield  {author} {\bibinfo {author} {\bibfnamefont {K.~D.}\ \bibnamefont
  {Irwin}},\ }\href {https://doi.org/10.1063/1.113674} {\bibfield  {journal}
  {\bibinfo  {journal} {Applied Physics Letters}\ }\textbf {\bibinfo {volume}
  {66}},\ \bibinfo {pages} {1998} (\bibinfo {year} {1995})}\BibitemShut
  {NoStop}%
\bibitem [{\citenamefont {Natarajan}\ \emph {et~al.}(2012)\citenamefont
  {Natarajan}, \citenamefont {Tanner},\ and\ \citenamefont
  {Hadfield}}]{Natarajan_2012}%
  \BibitemOpen
  \bibfield  {author} {\bibinfo {author} {\bibfnamefont {C.~M.}\ \bibnamefont
  {Natarajan}}, \bibinfo {author} {\bibfnamefont {M.~G.}\ \bibnamefont
  {Tanner}},\ and\ \bibinfo {author} {\bibfnamefont {R.~H.}\ \bibnamefont
  {Hadfield}},\ }\href {https://doi.org/10.1088/0953-2048/25/6/063001}
  {\bibfield  {journal} {\bibinfo  {journal} {Superconductor Science and
  Technology}\ }\textbf {\bibinfo {volume} {25}},\ \bibinfo {pages} {063001}
  (\bibinfo {year} {2012})}\BibitemShut {NoStop}%
\bibitem [{\citenamefont {Walls}\ and\ \citenamefont
  {Milburn}(2008)}]{walls2008quantum}%
  \BibitemOpen
  \bibfield  {author} {\bibinfo {author} {\bibfnamefont {D.~F.}\ \bibnamefont
  {Walls}}\ and\ \bibinfo {author} {\bibfnamefont {G.~J.}\ \bibnamefont
  {Milburn}},\ }\href@noop {} {\emph {\bibinfo {title} {Quantum optics}}}\
  (\bibinfo  {publisher} {Springer Science \& Business Media},\ \bibinfo {year}
  {2008})\BibitemShut {NoStop}%
\bibitem [{\citenamefont {Carusotto}\ and\ \citenamefont
  {Ciuti}(2013)}]{Carusotto_2013}%
  \BibitemOpen
  \bibfield  {author} {\bibinfo {author} {\bibfnamefont {I.}~\bibnamefont
  {Carusotto}}\ and\ \bibinfo {author} {\bibfnamefont {C.}~\bibnamefont
  {Ciuti}},\ }\href {https://doi.org/10.1103/RevModPhys.85.299} {\bibfield
  {journal} {\bibinfo  {journal} {Rev. Mod. Phys.}\ }\textbf {\bibinfo {volume}
  {85}},\ \bibinfo {pages} {299} (\bibinfo {year} {2013})}\BibitemShut
  {NoStop}%
\bibitem [{\citenamefont {Bloch}\ \emph {et~al.}(2022)\citenamefont {Bloch},
  \citenamefont {Carusotto},\ and\ \citenamefont {Wouters}}]{bloch2022non}%
  \BibitemOpen
  \bibfield  {author} {\bibinfo {author} {\bibfnamefont {J.}~\bibnamefont
  {Bloch}}, \bibinfo {author} {\bibfnamefont {I.}~\bibnamefont {Carusotto}},\
  and\ \bibinfo {author} {\bibfnamefont {M.}~\bibnamefont {Wouters}},\
  }\href@noop {} {\bibfield  {journal} {\bibinfo  {journal} {Nature Reviews
  Physics}\ }\textbf {\bibinfo {volume} {4}},\ \bibinfo {pages} {470} (\bibinfo
  {year} {2022})}\BibitemShut {NoStop}%
\bibitem [{\citenamefont {Graham}\ and\ \citenamefont
  {Haken}(1970)}]{graham1970laserlight}%
  \BibitemOpen
  \bibfield  {author} {\bibinfo {author} {\bibfnamefont {R.}~\bibnamefont
  {Graham}}\ and\ \bibinfo {author} {\bibfnamefont {H.}~\bibnamefont {Haken}},\
  }\href@noop {} {\bibfield  {journal} {\bibinfo  {journal} {Zeitschrift
  f{\"u}r Physik}\ }\textbf {\bibinfo {volume} {237}},\ \bibinfo {pages} {31}
  (\bibinfo {year} {1970})}\BibitemShut {NoStop}%
\bibitem [{\citenamefont {DeGiorgio}\ and\ \citenamefont
  {Scully}(1970)}]{Degiorgio_1970}%
  \BibitemOpen
  \bibfield  {author} {\bibinfo {author} {\bibfnamefont {V.}~\bibnamefont
  {DeGiorgio}}\ and\ \bibinfo {author} {\bibfnamefont {M.~O.}\ \bibnamefont
  {Scully}},\ }\href {https://doi.org/10.1103/PhysRevA.2.1170} {\bibfield
  {journal} {\bibinfo  {journal} {Phys. Rev. A}\ }\textbf {\bibinfo {volume}
  {2}},\ \bibinfo {pages} {1170} (\bibinfo {year} {1970})}\BibitemShut
  {NoStop}%
\bibitem [{\citenamefont {Gatti}\ and\ \citenamefont
  {Lugiato}(1995)}]{Gatti_1995}%
  \BibitemOpen
  \bibfield  {author} {\bibinfo {author} {\bibfnamefont {A.}~\bibnamefont
  {Gatti}}\ and\ \bibinfo {author} {\bibfnamefont {L.}~\bibnamefont
  {Lugiato}},\ }\href {https://doi.org/10.1103/PhysRevA.52.1675} {\bibfield
  {journal} {\bibinfo  {journal} {Phys. Rev. A}\ }\textbf {\bibinfo {volume}
  {52}},\ \bibinfo {pages} {1675} (\bibinfo {year} {1995})}\BibitemShut
  {NoStop}%
\bibitem [{\citenamefont {Bonifacio}\ and\ \citenamefont
  {Lugiato}(1978)}]{Bonifacio_1978}%
  \BibitemOpen
  \bibfield  {author} {\bibinfo {author} {\bibfnamefont {R.}~\bibnamefont
  {Bonifacio}}\ and\ \bibinfo {author} {\bibfnamefont {L.~A.}\ \bibnamefont
  {Lugiato}},\ }\href {https://doi.org/10.1103/PhysRevLett.40.1023} {\bibfield
  {journal} {\bibinfo  {journal} {Phys. Rev. Lett.}\ }\textbf {\bibinfo
  {volume} {40}},\ \bibinfo {pages} {1023} (\bibinfo {year}
  {1978})}\BibitemShut {NoStop}%
\bibitem [{\citenamefont {Drummond}\ and\ \citenamefont
  {Walls}(1980)}]{Drummond_1980}%
  \BibitemOpen
  \bibfield  {author} {\bibinfo {author} {\bibfnamefont {P.~D.}\ \bibnamefont
  {Drummond}}\ and\ \bibinfo {author} {\bibfnamefont {D.~F.}\ \bibnamefont
  {Walls}},\ }\href {https://doi.org/10.1088/0305-4470/13/2/034} {\bibfield
  {journal} {\bibinfo  {journal} {Journal of Physics A: Mathematical and
  General}\ }\textbf {\bibinfo {volume} {13}},\ \bibinfo {pages} {725}
  (\bibinfo {year} {1980})}\BibitemShut {NoStop}%
\bibitem [{\citenamefont {Bartolo}\ \emph {et~al.}(2016)\citenamefont
  {Bartolo}, \citenamefont {Minganti}, \citenamefont {Casteels},\ and\
  \citenamefont {Ciuti}}]{Ciuti_1and2}%
  \BibitemOpen
  \bibfield  {author} {\bibinfo {author} {\bibfnamefont {N.}~\bibnamefont
  {Bartolo}}, \bibinfo {author} {\bibfnamefont {F.}~\bibnamefont {Minganti}},
  \bibinfo {author} {\bibfnamefont {W.}~\bibnamefont {Casteels}},\ and\
  \bibinfo {author} {\bibfnamefont {C.}~\bibnamefont {Ciuti}},\ }\href
  {https://doi.org/10.1103/PhysRevA.94.033841} {\bibfield  {journal} {\bibinfo
  {journal} {Phys. Rev. A}\ }\textbf {\bibinfo {volume} {94}},\ \bibinfo
  {pages} {033841} (\bibinfo {year} {2016})}\BibitemShut {NoStop}%
\bibitem [{\citenamefont {Beaulieu}\ \emph
  {et~al.}(2025{\natexlab{a}})\citenamefont {Beaulieu}, \citenamefont
  {Minganti}, \citenamefont {Frasca}, \citenamefont {Savona}, \citenamefont
  {Felicetti}, \citenamefont {Di~Candia},\ and\ \citenamefont
  {Scarlino}}]{Beaulieu_2025}%
  \BibitemOpen
  \bibfield  {author} {\bibinfo {author} {\bibfnamefont {G.}~\bibnamefont
  {Beaulieu}}, \bibinfo {author} {\bibfnamefont {F.}~\bibnamefont {Minganti}},
  \bibinfo {author} {\bibfnamefont {S.}~\bibnamefont {Frasca}}, \bibinfo
  {author} {\bibfnamefont {V.}~\bibnamefont {Savona}}, \bibinfo {author}
  {\bibfnamefont {S.}~\bibnamefont {Felicetti}}, \bibinfo {author}
  {\bibfnamefont {R.}~\bibnamefont {Di~Candia}},\ and\ \bibinfo {author}
  {\bibfnamefont {P.}~\bibnamefont {Scarlino}},\ }\href
  {https://doi.org/10.1038/s41467-025-56830-w} {\bibfield  {journal} {\bibinfo
  {journal} {Nature Communications}\ }\textbf {\bibinfo {volume} {16}},\
  \bibinfo {pages} {1954} (\bibinfo {year} {2025}{\natexlab{a}})}\BibitemShut
  {NoStop}%
\bibitem [{\citenamefont {Casteels}\ \emph {et~al.}(2017)\citenamefont
  {Casteels}, \citenamefont {Fazio},\ and\ \citenamefont
  {Ciuti}}]{Ciuti_1pdrive}%
  \BibitemOpen
  \bibfield  {author} {\bibinfo {author} {\bibfnamefont {W.}~\bibnamefont
  {Casteels}}, \bibinfo {author} {\bibfnamefont {R.}~\bibnamefont {Fazio}},\
  and\ \bibinfo {author} {\bibfnamefont {C.}~\bibnamefont {Ciuti}},\ }\href
  {https://doi.org/10.1103/PhysRevA.95.012128} {\bibfield  {journal} {\bibinfo
  {journal} {Phys. Rev. A}\ }\textbf {\bibinfo {volume} {95}},\ \bibinfo
  {pages} {012128} (\bibinfo {year} {2017})}\BibitemShut {NoStop}%
\bibitem [{\citenamefont {Bakemeier}\ \emph {et~al.}(2012)\citenamefont
  {Bakemeier}, \citenamefont {Alvermann},\ and\ \citenamefont
  {Fehske}}]{Bakemeier_2012}%
  \BibitemOpen
  \bibfield  {author} {\bibinfo {author} {\bibfnamefont {L.}~\bibnamefont
  {Bakemeier}}, \bibinfo {author} {\bibfnamefont {A.}~\bibnamefont
  {Alvermann}},\ and\ \bibinfo {author} {\bibfnamefont {H.}~\bibnamefont
  {Fehske}},\ }\href {https://doi.org/10.1103/PhysRevA.85.043821} {\bibfield
  {journal} {\bibinfo  {journal} {Phys. Rev. A}\ }\textbf {\bibinfo {volume}
  {85}},\ \bibinfo {pages} {043821} (\bibinfo {year} {2012})}\BibitemShut
  {NoStop}%
\bibitem [{\citenamefont {Puebla}\ \emph {et~al.}(2017)\citenamefont {Puebla},
  \citenamefont {Hwang}, \citenamefont {Casanova},\ and\ \citenamefont
  {Plenio}}]{Ricardo_2017}%
  \BibitemOpen
  \bibfield  {author} {\bibinfo {author} {\bibfnamefont {R.}~\bibnamefont
  {Puebla}}, \bibinfo {author} {\bibfnamefont {M.-J.}\ \bibnamefont {Hwang}},
  \bibinfo {author} {\bibfnamefont {J.}~\bibnamefont {Casanova}},\ and\
  \bibinfo {author} {\bibfnamefont {M.~B.}\ \bibnamefont {Plenio}},\ }\href
  {https://doi.org/10.1103/PhysRevLett.118.073001} {\bibfield  {journal}
  {\bibinfo  {journal} {Phys. Rev. Lett.}\ }\textbf {\bibinfo {volume} {118}},\
  \bibinfo {pages} {073001} (\bibinfo {year} {2017})}\BibitemShut {NoStop}%
\bibitem [{\citenamefont {Greentree}\ \emph {et~al.}(2006)\citenamefont
  {Greentree}, \citenamefont {Tahan}, \citenamefont {Cole},\ and\ \citenamefont
  {Hollenberg}}]{Greentree_2006}%
  \BibitemOpen
  \bibfield  {author} {\bibinfo {author} {\bibfnamefont {A.~D.}\ \bibnamefont
  {Greentree}}, \bibinfo {author} {\bibfnamefont {C.}~\bibnamefont {Tahan}},
  \bibinfo {author} {\bibfnamefont {J.~H.}\ \bibnamefont {Cole}},\ and\
  \bibinfo {author} {\bibfnamefont {L.~C.~L.}\ \bibnamefont {Hollenberg}},\
  }\href {https://doi.org/10.1038/nphys466} {\bibfield  {journal} {\bibinfo
  {journal} {Nature Physics}\ }\textbf {\bibinfo {volume} {2}},\ \bibinfo
  {pages} {856} (\bibinfo {year} {2006})}\BibitemShut {NoStop}%
\bibitem [{\citenamefont {Angelakis}\ \emph {et~al.}(2007)\citenamefont
  {Angelakis}, \citenamefont {Santos},\ and\ \citenamefont
  {Bose}}]{Angelakis:PRA2007}%
  \BibitemOpen
  \bibfield  {author} {\bibinfo {author} {\bibfnamefont {D.~G.}\ \bibnamefont
  {Angelakis}}, \bibinfo {author} {\bibfnamefont {M.~F.}\ \bibnamefont
  {Santos}},\ and\ \bibinfo {author} {\bibfnamefont {S.}~\bibnamefont {Bose}},\
  }\bibfield  {journal} {\bibinfo  {journal} {Phys. Rev. A}\ }\textbf {\bibinfo
  {volume} {76}},\ \href {https://doi.org/10.1103/PhysRevA.76.031805}
  {10.1103/PhysRevA.76.031805} (\bibinfo {year} {2007})\BibitemShut {NoStop}%
\bibitem [{\citenamefont {Hartmann}\ \emph {et~al.}(2006)\citenamefont
  {Hartmann}, \citenamefont {Brand\~ao},\ and\ \citenamefont
  {Plenio}}]{Hartmann:NatPhys2006}%
  \BibitemOpen
  \bibfield  {author} {\bibinfo {author} {\bibfnamefont {M.~J.}\ \bibnamefont
  {Hartmann}}, \bibinfo {author} {\bibfnamefont {F.~G. S.~L.}\ \bibnamefont
  {Brand\~ao}},\ and\ \bibinfo {author} {\bibfnamefont {M.~B.}\ \bibnamefont
  {Plenio}},\ }\href {https://doi.org/10.1038/nphys462} {\bibfield  {journal}
  {\bibinfo  {journal} {Nature Phys.}\ }\textbf {\bibinfo {volume} {2}},\
  \bibinfo {pages} {849} (\bibinfo {year} {2006})}\BibitemShut {NoStop}%
\bibitem [{\citenamefont {Lebreuilly}\ \emph {et~al.}(2017)\citenamefont
  {Lebreuilly}, \citenamefont {Biella}, \citenamefont {Storme}, \citenamefont
  {Rossini}, \citenamefont {Fazio}, \citenamefont {Ciuti},\ and\ \citenamefont
  {Carusotto}}]{lebreuilly2017stabilizing}%
  \BibitemOpen
  \bibfield  {author} {\bibinfo {author} {\bibfnamefont {J.}~\bibnamefont
  {Lebreuilly}}, \bibinfo {author} {\bibfnamefont {A.}~\bibnamefont {Biella}},
  \bibinfo {author} {\bibfnamefont {F.}~\bibnamefont {Storme}}, \bibinfo
  {author} {\bibfnamefont {D.}~\bibnamefont {Rossini}}, \bibinfo {author}
  {\bibfnamefont {R.}~\bibnamefont {Fazio}}, \bibinfo {author} {\bibfnamefont
  {C.}~\bibnamefont {Ciuti}},\ and\ \bibinfo {author} {\bibfnamefont
  {I.}~\bibnamefont {Carusotto}},\ }\href@noop {} {\bibfield  {journal}
  {\bibinfo  {journal} {Physical Review A}\ }\textbf {\bibinfo {volume} {96}},\
  \bibinfo {pages} {033828} (\bibinfo {year} {2017})}\BibitemShut {NoStop}%
\bibitem [{\citenamefont {Biella}\ \emph {et~al.}(2017)\citenamefont {Biella},
  \citenamefont {Storme}, \citenamefont {Lebreuilly}, \citenamefont {Rossini},
  \citenamefont {Fazio}, \citenamefont {Carusotto},\ and\ \citenamefont
  {Ciuti}}]{biella2017phase}%
  \BibitemOpen
  \bibfield  {author} {\bibinfo {author} {\bibfnamefont {A.}~\bibnamefont
  {Biella}}, \bibinfo {author} {\bibfnamefont {F.}~\bibnamefont {Storme}},
  \bibinfo {author} {\bibfnamefont {J.}~\bibnamefont {Lebreuilly}}, \bibinfo
  {author} {\bibfnamefont {D.}~\bibnamefont {Rossini}}, \bibinfo {author}
  {\bibfnamefont {R.}~\bibnamefont {Fazio}}, \bibinfo {author} {\bibfnamefont
  {I.}~\bibnamefont {Carusotto}},\ and\ \bibinfo {author} {\bibfnamefont
  {C.}~\bibnamefont {Ciuti}},\ }\href@noop {} {\bibfield  {journal} {\bibinfo
  {journal} {Physical Review A}\ }\textbf {\bibinfo {volume} {96}},\ \bibinfo
  {pages} {023839} (\bibinfo {year} {2017})}\BibitemShut {NoStop}%
\bibitem [{\citenamefont {Carmichael}(2015)}]{Carmichael_2015}%
  \BibitemOpen
  \bibfield  {author} {\bibinfo {author} {\bibfnamefont {H.~J.}\ \bibnamefont
  {Carmichael}},\ }\href {https://doi.org/10.1103/PhysRevX.5.031028} {\bibfield
   {journal} {\bibinfo  {journal} {Phys. Rev. X}\ }\textbf {\bibinfo {volume}
  {5}},\ \bibinfo {pages} {031028} (\bibinfo {year} {2015})}\BibitemShut
  {NoStop}%
\bibitem [{\citenamefont {Ma}\ \emph {et~al.}(2019)\citenamefont {Ma},
  \citenamefont {Saxberg}, \citenamefont {Owens}, \citenamefont {Leung},
  \citenamefont {Lu}, \citenamefont {Simon},\ and\ \citenamefont
  {Schuster}}]{ma2019dissipatively}%
  \BibitemOpen
  \bibfield  {author} {\bibinfo {author} {\bibfnamefont {R.}~\bibnamefont
  {Ma}}, \bibinfo {author} {\bibfnamefont {B.}~\bibnamefont {Saxberg}},
  \bibinfo {author} {\bibfnamefont {C.}~\bibnamefont {Owens}}, \bibinfo
  {author} {\bibfnamefont {N.}~\bibnamefont {Leung}}, \bibinfo {author}
  {\bibfnamefont {Y.}~\bibnamefont {Lu}}, \bibinfo {author} {\bibfnamefont
  {J.}~\bibnamefont {Simon}},\ and\ \bibinfo {author} {\bibfnamefont {D.~I.}\
  \bibnamefont {Schuster}},\ }\href@noop {} {\bibfield  {journal} {\bibinfo
  {journal} {Nature}\ }\textbf {\bibinfo {volume} {566}},\ \bibinfo {pages}
  {51} (\bibinfo {year} {2019})}\BibitemShut {NoStop}%
\bibitem [{\citenamefont {Wiesenfeld}\ and\ \citenamefont
  {McNamara}(1985)}]{Wiesenfeld:PRL1985}%
  \BibitemOpen
  \bibfield  {author} {\bibinfo {author} {\bibfnamefont {K.}~\bibnamefont
  {Wiesenfeld}}\ and\ \bibinfo {author} {\bibfnamefont {B.}~\bibnamefont
  {McNamara}},\ }\href {https://doi.org/10.1103/PhysRevLett.55.13} {\bibfield
  {journal} {\bibinfo  {journal} {Phys. Rev. Lett.}\ }\textbf {\bibinfo
  {volume} {55}},\ \bibinfo {pages} {13} (\bibinfo {year} {1985})}\BibitemShut
  {NoStop}%
\bibitem [{\citenamefont {Vijay}\ \emph {et~al.}(2009)\citenamefont {Vijay},
  \citenamefont {Devoret},\ and\ \citenamefont {Siddiqi}}]{Vijay_2009}%
  \BibitemOpen
  \bibfield  {author} {\bibinfo {author} {\bibfnamefont {R.}~\bibnamefont
  {Vijay}}, \bibinfo {author} {\bibfnamefont {M.~H.}\ \bibnamefont {Devoret}},\
  and\ \bibinfo {author} {\bibfnamefont {I.}~\bibnamefont {Siddiqi}},\ }\href
  {https://doi.org/10.1063/1.3224703} {\bibfield  {journal} {\bibinfo
  {journal} {Review of Scientific Instruments}\ }\textbf {\bibinfo {volume}
  {80}},\ \bibinfo {pages} {111101} (\bibinfo {year} {2009})}\BibitemShut
  {NoStop}%
\bibitem [{\citenamefont {Zanardi}\ \emph {et~al.}(2008)\citenamefont
  {Zanardi}, \citenamefont {Paris},\ and\ \citenamefont
  {Campos~Venuti}}]{Zanardi:PRA2008}%
  \BibitemOpen
  \bibfield  {author} {\bibinfo {author} {\bibfnamefont {P.}~\bibnamefont
  {Zanardi}}, \bibinfo {author} {\bibfnamefont {M.~G.~A.}\ \bibnamefont
  {Paris}},\ and\ \bibinfo {author} {\bibfnamefont {L.}~\bibnamefont
  {Campos~Venuti}},\ }\href {https://doi.org/10.1103/PhysRevA.78.042105}
  {\bibfield  {journal} {\bibinfo  {journal} {Phys. Rev. A}\ }\textbf {\bibinfo
  {volume} {78}},\ \bibinfo {pages} {042105} (\bibinfo {year}
  {2008})}\BibitemShut {NoStop}%
\bibitem [{\citenamefont {Garbe}\ \emph {et~al.}(2020)\citenamefont {Garbe},
  \citenamefont {Bina}, \citenamefont {Keller}, \citenamefont {Paris},\ and\
  \citenamefont {Felicetti}}]{Garbe:PRL2020}%
  \BibitemOpen
  \bibfield  {author} {\bibinfo {author} {\bibfnamefont {L.}~\bibnamefont
  {Garbe}}, \bibinfo {author} {\bibfnamefont {M.}~\bibnamefont {Bina}},
  \bibinfo {author} {\bibfnamefont {A.}~\bibnamefont {Keller}}, \bibinfo
  {author} {\bibfnamefont {M.~G.~A.}\ \bibnamefont {Paris}},\ and\ \bibinfo
  {author} {\bibfnamefont {S.}~\bibnamefont {Felicetti}},\ }\href
  {https://doi.org/10.1103/PhysRevLett.124.120504} {\bibfield  {journal}
  {\bibinfo  {journal} {Phys. Rev. Lett.}\ }\textbf {\bibinfo {volume} {124}},\
  \bibinfo {pages} {120504} (\bibinfo {year} {2020})}\BibitemShut {NoStop}%
\bibitem [{\citenamefont {Roy}\ \emph {et~al.}(2021)\citenamefont {Roy},
  \citenamefont {Jahani}, \citenamefont {Langrock}, \citenamefont {Fejer},\
  and\ \citenamefont {Marandi}}]{Roy_2021}%
  \BibitemOpen
  \bibfield  {author} {\bibinfo {author} {\bibfnamefont {A.}~\bibnamefont
  {Roy}}, \bibinfo {author} {\bibfnamefont {S.}~\bibnamefont {Jahani}},
  \bibinfo {author} {\bibfnamefont {C.}~\bibnamefont {Langrock}}, \bibinfo
  {author} {\bibfnamefont {M.}~\bibnamefont {Fejer}},\ and\ \bibinfo {author}
  {\bibfnamefont {A.}~\bibnamefont {Marandi}},\ }\href
  {https://doi.org/10.1038/s41467-021-21048-z} {\bibfield  {journal} {\bibinfo
  {journal} {Nature Communications}\ }\textbf {\bibinfo {volume} {12}},\
  \bibinfo {pages} {835} (\bibinfo {year} {2021})}\BibitemShut {NoStop}%
\bibitem [{\citenamefont {Di~Candia}\ \emph {et~al.}(2023)\citenamefont
  {Di~Candia}, \citenamefont {Minganti}, \citenamefont {Petrovnin},
  \citenamefont {Paraoanu},\ and\ \citenamefont
  {Felicetti}}]{DiCandia:NPJQI2023}%
  \BibitemOpen
  \bibfield  {author} {\bibinfo {author} {\bibfnamefont {R.}~\bibnamefont
  {Di~Candia}}, \bibinfo {author} {\bibfnamefont {F.}~\bibnamefont {Minganti}},
  \bibinfo {author} {\bibfnamefont {K.}~\bibnamefont {Petrovnin}}, \bibinfo
  {author} {\bibfnamefont {G.~S.}\ \bibnamefont {Paraoanu}},\ and\ \bibinfo
  {author} {\bibfnamefont {S.}~\bibnamefont {Felicetti}},\ }\href@noop {}
  {\bibfield  {journal} {\bibinfo  {journal} {npj Quantum Information}\
  }\textbf {\bibinfo {volume} {9}},\ \bibinfo {pages} {23} (\bibinfo {year}
  {2023})}\BibitemShut {NoStop}%
\bibitem [{\citenamefont {Beaulieu}\ \emph
  {et~al.}(2025{\natexlab{b}})\citenamefont {Beaulieu}, \citenamefont
  {Minganti}, \citenamefont {Frasca}, \citenamefont {Scigliuzzo}, \citenamefont
  {Felicetti}, \citenamefont {Di~Candia},\ and\ \citenamefont
  {Scarlino}}]{Beaulieu:PRXQ2025}%
  \BibitemOpen
  \bibfield  {author} {\bibinfo {author} {\bibfnamefont {G.}~\bibnamefont
  {Beaulieu}}, \bibinfo {author} {\bibfnamefont {F.}~\bibnamefont {Minganti}},
  \bibinfo {author} {\bibfnamefont {S.}~\bibnamefont {Frasca}}, \bibinfo
  {author} {\bibfnamefont {M.}~\bibnamefont {Scigliuzzo}}, \bibinfo {author}
  {\bibfnamefont {S.}~\bibnamefont {Felicetti}}, \bibinfo {author}
  {\bibfnamefont {R.}~\bibnamefont {Di~Candia}},\ and\ \bibinfo {author}
  {\bibfnamefont {P.}~\bibnamefont {Scarlino}},\ }\href
  {https://doi.org/10.1103/PRXQuantum.6.020301} {\bibfield  {journal} {\bibinfo
   {journal} {PRX Quantum}\ }\textbf {\bibinfo {volume} {6}},\ \bibinfo {pages}
  {020301} (\bibinfo {year} {2025}{\natexlab{b}})}\BibitemShut {NoStop}%
\bibitem [{\citenamefont {Tang}\ \emph {et~al.}(2023)\citenamefont {Tang},
  \citenamefont {Qin}, \citenamefont {Liu}, \citenamefont {Wang}, \citenamefont
  {Cui}, \citenamefont {Su}, \citenamefont {Yan},\ and\ \citenamefont
  {Chen}}]{Tang:PRA2023}%
  \BibitemOpen
  \bibfield  {author} {\bibinfo {author} {\bibfnamefont {S.-B.}\ \bibnamefont
  {Tang}}, \bibinfo {author} {\bibfnamefont {H.}~\bibnamefont {Qin}}, \bibinfo
  {author} {\bibfnamefont {B.-B.}\ \bibnamefont {Liu}}, \bibinfo {author}
  {\bibfnamefont {D.-Y.}\ \bibnamefont {Wang}}, \bibinfo {author}
  {\bibfnamefont {K.}~\bibnamefont {Cui}}, \bibinfo {author} {\bibfnamefont
  {S.-L.}\ \bibnamefont {Su}}, \bibinfo {author} {\bibfnamefont {L.-L.}\
  \bibnamefont {Yan}},\ and\ \bibinfo {author} {\bibfnamefont {G.}~\bibnamefont
  {Chen}},\ }\href {https://doi.org/10.1103/PhysRevA.108.053514} {\bibfield
  {journal} {\bibinfo  {journal} {Phys. Rev. A}\ }\textbf {\bibinfo {volume}
  {108}},\ \bibinfo {pages} {053514} (\bibinfo {year} {2023})}\BibitemShut
  {NoStop}%
\bibitem [{\citenamefont {Petrovnin}\ \emph {et~al.}(2024)\citenamefont
  {Petrovnin}, \citenamefont {Wang}, \citenamefont {Perelshtein}, \citenamefont
  {Hakonen},\ and\ \citenamefont {Paraoanu}}]{Petrovnin:PRXQ2024}%
  \BibitemOpen
  \bibfield  {author} {\bibinfo {author} {\bibfnamefont {K.}~\bibnamefont
  {Petrovnin}}, \bibinfo {author} {\bibfnamefont {J.}~\bibnamefont {Wang}},
  \bibinfo {author} {\bibfnamefont {M.}~\bibnamefont {Perelshtein}}, \bibinfo
  {author} {\bibfnamefont {P.}~\bibnamefont {Hakonen}},\ and\ \bibinfo {author}
  {\bibfnamefont {G.~S.}\ \bibnamefont {Paraoanu}},\ }\href
  {https://doi.org/10.1103/PRXQuantum.5.020342} {\bibfield  {journal} {\bibinfo
   {journal} {PRX Quantum}\ }\textbf {\bibinfo {volume} {5}},\ \bibinfo {pages}
  {020342} (\bibinfo {year} {2024})}\BibitemShut {NoStop}%
\bibitem [{\citenamefont {Vogel}\ and\ \citenamefont
  {Risken}(1988)}]{vogel1988quantum}%
  \BibitemOpen
  \bibfield  {author} {\bibinfo {author} {\bibfnamefont {K.}~\bibnamefont
  {Vogel}}\ and\ \bibinfo {author} {\bibfnamefont {H.}~\bibnamefont {Risken}},\
  }\href@noop {} {\bibfield  {journal} {\bibinfo  {journal} {Physical Review
  A}\ }\textbf {\bibinfo {volume} {38}},\ \bibinfo {pages} {2409} (\bibinfo
  {year} {1988})}\BibitemShut {NoStop}%
\bibitem [{\citenamefont {Vogel}\ and\ \citenamefont
  {Risken}(1989)}]{vogel1989quasiprobability}%
  \BibitemOpen
  \bibfield  {author} {\bibinfo {author} {\bibfnamefont {K.}~\bibnamefont
  {Vogel}}\ and\ \bibinfo {author} {\bibfnamefont {H.}~\bibnamefont {Risken}},\
  }\href@noop {} {\bibfield  {journal} {\bibinfo  {journal} {Physical Review
  A}\ }\textbf {\bibinfo {volume} {39}},\ \bibinfo {pages} {4675} (\bibinfo
  {year} {1989})}\BibitemShut {NoStop}%
\bibitem [{\citenamefont {Rodriguez}\ \emph {et~al.}(2017)\citenamefont
  {Rodriguez}, \citenamefont {Casteels}, \citenamefont {Storme}, \citenamefont
  {Carlon~Zambon}, \citenamefont {Sagnes}, \citenamefont {Le~Gratiet},
  \citenamefont {Galopin}, \citenamefont {Lema\^itre}, \citenamefont {Amo},
  \citenamefont {Ciuti},\ and\ \citenamefont {Bloch}}]{Rodriguez_2017}%
  \BibitemOpen
  \bibfield  {author} {\bibinfo {author} {\bibfnamefont {S.~R.~K.}\
  \bibnamefont {Rodriguez}}, \bibinfo {author} {\bibfnamefont {W.}~\bibnamefont
  {Casteels}}, \bibinfo {author} {\bibfnamefont {F.}~\bibnamefont {Storme}},
  \bibinfo {author} {\bibfnamefont {N.}~\bibnamefont {Carlon~Zambon}}, \bibinfo
  {author} {\bibfnamefont {I.}~\bibnamefont {Sagnes}}, \bibinfo {author}
  {\bibfnamefont {L.}~\bibnamefont {Le~Gratiet}}, \bibinfo {author}
  {\bibfnamefont {E.}~\bibnamefont {Galopin}}, \bibinfo {author} {\bibfnamefont
  {A.}~\bibnamefont {Lema\^itre}}, \bibinfo {author} {\bibfnamefont
  {A.}~\bibnamefont {Amo}}, \bibinfo {author} {\bibfnamefont {C.}~\bibnamefont
  {Ciuti}},\ and\ \bibinfo {author} {\bibfnamefont {J.}~\bibnamefont {Bloch}},\
  }\href {https://doi.org/10.1103/PhysRevLett.118.247402} {\bibfield  {journal}
  {\bibinfo  {journal} {Physical Review Letters}\ }\textbf {\bibinfo {volume}
  {118}},\ \bibinfo {pages} {247402} (\bibinfo {year} {2017})}\BibitemShut
  {NoStop}%
\bibitem [{\citenamefont {Fitzpatrick}\ \emph {et~al.}(2017)\citenamefont
  {Fitzpatrick}, \citenamefont {Sundaresan}, \citenamefont {Li}, \citenamefont
  {Koch},\ and\ \citenamefont {Houck}}]{Fitz_2017}%
  \BibitemOpen
  \bibfield  {author} {\bibinfo {author} {\bibfnamefont {M.}~\bibnamefont
  {Fitzpatrick}}, \bibinfo {author} {\bibfnamefont {N.~M.}\ \bibnamefont
  {Sundaresan}}, \bibinfo {author} {\bibfnamefont {A.~C.~Y.}\ \bibnamefont
  {Li}}, \bibinfo {author} {\bibfnamefont {J.}~\bibnamefont {Koch}},\ and\
  \bibinfo {author} {\bibfnamefont {A.~A.}\ \bibnamefont {Houck}},\ }\href
  {https://doi.org/10.1103/PhysRevX.7.011016} {\bibfield  {journal} {\bibinfo
  {journal} {Phys. Rev. X}\ }\textbf {\bibinfo {volume} {7}},\ \bibinfo {pages}
  {011016} (\bibinfo {year} {2017})}\BibitemShut {NoStop}%
\bibitem [{\citenamefont {Fink}\ \emph {et~al.}(2018)\citenamefont {Fink},
  \citenamefont {Schade}, \citenamefont {H{\"o}fling}, \citenamefont
  {Schneider},\ and\ \citenamefont {Imamoglu}}]{Fink2018}%
  \BibitemOpen
  \bibfield  {author} {\bibinfo {author} {\bibfnamefont {T.}~\bibnamefont
  {Fink}}, \bibinfo {author} {\bibfnamefont {A.}~\bibnamefont {Schade}},
  \bibinfo {author} {\bibfnamefont {S.}~\bibnamefont {H{\"o}fling}}, \bibinfo
  {author} {\bibfnamefont {C.}~\bibnamefont {Schneider}},\ and\ \bibinfo
  {author} {\bibfnamefont {A.}~\bibnamefont {Imamoglu}},\ }\href
  {https://doi.org/10.1038/s41567-017-0020-9} {\bibfield  {journal} {\bibinfo
  {journal} {Nature Physics}\ }\textbf {\bibinfo {volume} {14}},\ \bibinfo
  {pages} {365} (\bibinfo {year} {2018})}\BibitemShut {NoStop}%
\bibitem [{\citenamefont
  {Carmichael}(1993{\natexlab{a}})}]{Carmicheal:PRL1993}%
  \BibitemOpen
  \bibfield  {author} {\bibinfo {author} {\bibfnamefont {H.~J.}\ \bibnamefont
  {Carmichael}},\ }\href {https://doi.org/10.1103/PhysRevLett.70.2273}
  {\bibfield  {journal} {\bibinfo  {journal} {Phys. Rev. Lett.}\ }\textbf
  {\bibinfo {volume} {70}},\ \bibinfo {pages} {2273} (\bibinfo {year}
  {1993}{\natexlab{a}})}\BibitemShut {NoStop}%
\bibitem [{\citenamefont {Breuer}\ and\ \citenamefont
  {Petruccione}(2002)}]{breuer2002theory}%
  \BibitemOpen
  \bibfield  {author} {\bibinfo {author} {\bibfnamefont {H.-P.}\ \bibnamefont
  {Breuer}}\ and\ \bibinfo {author} {\bibfnamefont {F.}~\bibnamefont
  {Petruccione}},\ }\href@noop {} {\emph {\bibinfo {title} {The theory of open
  quantum systems}}}\ (\bibinfo  {publisher} {OUP Oxford},\ \bibinfo {year}
  {2002})\BibitemShut {NoStop}%
\bibitem [{\citenamefont
  {Carmichael}(1993{\natexlab{b}})}]{carmichael1993open}%
  \BibitemOpen
  \bibfield  {author} {\bibinfo {author} {\bibfnamefont {H.}~\bibnamefont
  {Carmichael}},\ }\href@noop {} {\emph {\bibinfo {title} {An open systems
  approach to quantum optics}}}\ (\bibinfo  {publisher} {Springer},\ \bibinfo
  {year} {1993})\BibitemShut {NoStop}%
\bibitem [{\citenamefont {Gardiner}(1993)}]{Gardiner:PRL1993}%
  \BibitemOpen
  \bibfield  {author} {\bibinfo {author} {\bibfnamefont {C.}~\bibnamefont
  {Gardiner}},\ }\href@noop {} {\bibfield  {journal} {\bibinfo  {journal}
  {Physical review letters}\ }\textbf {\bibinfo {volume} {70}},\ \bibinfo
  {pages} {2269} (\bibinfo {year} {1993})}\BibitemShut {NoStop}%
\bibitem [{\citenamefont {Johansson}\ \emph {et~al.}(2012)\citenamefont
  {Johansson}, \citenamefont {Nation},\ and\ \citenamefont
  {Nori}}]{johansson2012qutip}%
  \BibitemOpen
  \bibfield  {author} {\bibinfo {author} {\bibfnamefont {J.}~\bibnamefont
  {Johansson}}, \bibinfo {author} {\bibfnamefont {P.}~\bibnamefont {Nation}},\
  and\ \bibinfo {author} {\bibfnamefont {F.}~\bibnamefont {Nori}},\ }\href
  {https://doi.org/https://doi.org/10.1016/j.cpc.2012.02.021} {\bibfield
  {journal} {\bibinfo  {journal} {Computer Physics Communications}\ }\textbf
  {\bibinfo {volume} {183}},\ \bibinfo {pages} {1760} (\bibinfo {year}
  {2012})}\BibitemShut {NoStop}%
\bibitem [{\citenamefont {Johansson}\ \emph {et~al.}(2013)\citenamefont
  {Johansson}, \citenamefont {Nation},\ and\ \citenamefont
  {Nori}}]{johansson2013qutip}%
  \BibitemOpen
  \bibfield  {author} {\bibinfo {author} {\bibfnamefont {J.}~\bibnamefont
  {Johansson}}, \bibinfo {author} {\bibfnamefont {P.}~\bibnamefont {Nation}},\
  and\ \bibinfo {author} {\bibfnamefont {F.}~\bibnamefont {Nori}},\ }\href
  {https://doi.org/https://doi.org/10.1016/j.cpc.2012.11.019} {\bibfield
  {journal} {\bibinfo  {journal} {Computer Physics Communications}\ }\textbf
  {\bibinfo {volume} {184}},\ \bibinfo {pages} {1234} (\bibinfo {year}
  {2013})}\BibitemShut {NoStop}%
\bibitem [{\citenamefont {Lambert}\ \emph {et~al.}(2026)\citenamefont
  {Lambert}, \citenamefont {Gigu\`ere}, \citenamefont {Menczel}, \citenamefont
  {Li}, \citenamefont {Hopf}, \citenamefont {Su\'arez}, \citenamefont {Gali},
  \citenamefont {Lishman}, \citenamefont {Gadhvi}, \citenamefont {Agarwal},
  \citenamefont {Galicia}, \citenamefont {Shammah}, \citenamefont {Nation},
  \citenamefont {Johansson}, \citenamefont {Ahmed}, \citenamefont {Cross},
  \citenamefont {Pitchford},\ and\ \citenamefont {Nori}}]{qutip5}%
  \BibitemOpen
  \bibfield  {author} {\bibinfo {author} {\bibfnamefont {N.}~\bibnamefont
  {Lambert}}, \bibinfo {author} {\bibfnamefont {E.}~\bibnamefont {Gigu\`ere}},
  \bibinfo {author} {\bibfnamefont {P.}~\bibnamefont {Menczel}}, \bibinfo
  {author} {\bibfnamefont {B.}~\bibnamefont {Li}}, \bibinfo {author}
  {\bibfnamefont {P.}~\bibnamefont {Hopf}}, \bibinfo {author} {\bibfnamefont
  {G.}~\bibnamefont {Su\'arez}}, \bibinfo {author} {\bibfnamefont
  {M.}~\bibnamefont {Gali}}, \bibinfo {author} {\bibfnamefont {J.}~\bibnamefont
  {Lishman}}, \bibinfo {author} {\bibfnamefont {R.}~\bibnamefont {Gadhvi}},
  \bibinfo {author} {\bibfnamefont {R.}~\bibnamefont {Agarwal}}, \bibinfo
  {author} {\bibfnamefont {A.}~\bibnamefont {Galicia}}, \bibinfo {author}
  {\bibfnamefont {N.}~\bibnamefont {Shammah}}, \bibinfo {author} {\bibfnamefont
  {P.}~\bibnamefont {Nation}}, \bibinfo {author} {\bibfnamefont {J.~R.}\
  \bibnamefont {Johansson}}, \bibinfo {author} {\bibfnamefont {S.}~\bibnamefont
  {Ahmed}}, \bibinfo {author} {\bibfnamefont {S.}~\bibnamefont {Cross}},
  \bibinfo {author} {\bibfnamefont {A.}~\bibnamefont {Pitchford}},\ and\
  \bibinfo {author} {\bibfnamefont {F.}~\bibnamefont {Nori}},\ }\href
  {https://doi.org/10.1016/j.physrep.2025.10.001} {\bibfield  {journal}
  {\bibinfo  {journal} {Physics Reports}\ }\textbf {\bibinfo {volume} {1153}},\
  \bibinfo {pages} {1} (\bibinfo {year} {2026})}\BibitemShut {NoStop}%
\bibitem [{\citenamefont {Boyd}()}]{boyd2003nonlinear}%
  \BibitemOpen
  \bibfield  {author} {\bibinfo {author} {\bibfnamefont {R.~W.}\ \bibnamefont
  {Boyd}},\ }\href@noop {} {\emph {\bibinfo {title} {Nonlinear Optics (2)}}},\
  Vol.\ \bibinfo {volume} {611}\BibitemShut {NoStop}%
\bibitem [{\citenamefont {Claude}\ \emph {et~al.}(2022)\citenamefont {Claude},
  \citenamefont {Jacquet}, \citenamefont {Usciati}, \citenamefont {Carusotto},
  \citenamefont {Giacobino}, \citenamefont {Bramati},\ and\ \citenamefont
  {Glorieux}}]{Claude:PRL2022}%
  \BibitemOpen
  \bibfield  {author} {\bibinfo {author} {\bibfnamefont {F.}~\bibnamefont
  {Claude}}, \bibinfo {author} {\bibfnamefont {M.~J.}\ \bibnamefont {Jacquet}},
  \bibinfo {author} {\bibfnamefont {R.}~\bibnamefont {Usciati}}, \bibinfo
  {author} {\bibfnamefont {I.}~\bibnamefont {Carusotto}}, \bibinfo {author}
  {\bibfnamefont {E.}~\bibnamefont {Giacobino}}, \bibinfo {author}
  {\bibfnamefont {A.}~\bibnamefont {Bramati}},\ and\ \bibinfo {author}
  {\bibfnamefont {Q.}~\bibnamefont {Glorieux}},\ }\href@noop {} {\bibfield
  {journal} {\bibinfo  {journal} {Physical Review Letters}\ }\textbf {\bibinfo
  {volume} {129}},\ \bibinfo {pages} {103601} (\bibinfo {year}
  {2022})}\BibitemShut {NoStop}%
\bibitem [{\citenamefont {Baas}\ \emph {et~al.}(2004)\citenamefont {Baas},
  \citenamefont {Karr}, \citenamefont {Eleuch},\ and\ \citenamefont
  {Giacobino}}]{PhysRevA.69.023809}%
  \BibitemOpen
  \bibfield  {author} {\bibinfo {author} {\bibfnamefont {A.}~\bibnamefont
  {Baas}}, \bibinfo {author} {\bibfnamefont {J.~P.}\ \bibnamefont {Karr}},
  \bibinfo {author} {\bibfnamefont {H.}~\bibnamefont {Eleuch}},\ and\ \bibinfo
  {author} {\bibfnamefont {E.}~\bibnamefont {Giacobino}},\ }\href
  {https://doi.org/10.1103/PhysRevA.69.023809} {\bibfield  {journal} {\bibinfo
  {journal} {Phys. Rev. A}\ }\textbf {\bibinfo {volume} {69}},\ \bibinfo
  {pages} {023809} (\bibinfo {year} {2004})}\BibitemShut {NoStop}%
\bibitem [{\citenamefont {Togan}\ \emph {et~al.}(2018)\citenamefont {Togan},
  \citenamefont {Lim}, \citenamefont {Faelt}, \citenamefont {Wegscheider},\
  and\ \citenamefont {Imamoglu}}]{Togan_2018}%
  \BibitemOpen
  \bibfield  {author} {\bibinfo {author} {\bibfnamefont {E.}~\bibnamefont
  {Togan}}, \bibinfo {author} {\bibfnamefont {H.-T.}\ \bibnamefont {Lim}},
  \bibinfo {author} {\bibfnamefont {S.}~\bibnamefont {Faelt}}, \bibinfo
  {author} {\bibfnamefont {W.}~\bibnamefont {Wegscheider}},\ and\ \bibinfo
  {author} {\bibfnamefont {A.}~\bibnamefont {Imamoglu}},\ }\href
  {https://doi.org/10.1103/PhysRevLett.121.227402} {\bibfield  {journal}
  {\bibinfo  {journal} {Phys. Rev. Lett.}\ }\textbf {\bibinfo {volume} {121}},\
  \bibinfo {pages} {227402} (\bibinfo {year} {2018})}\BibitemShut {NoStop}%
\bibitem [{\citenamefont {Rosenberg}\ \emph {et~al.}(2018)\citenamefont
  {Rosenberg}, \citenamefont {Liran}, \citenamefont {Mazuz-Harpaz},
  \citenamefont {West}, \citenamefont {Pfeiffer},\ and\ \citenamefont
  {Rapaport}}]{Rosenberg_2018}%
  \BibitemOpen
  \bibfield  {author} {\bibinfo {author} {\bibfnamefont {I.}~\bibnamefont
  {Rosenberg}}, \bibinfo {author} {\bibfnamefont {D.}~\bibnamefont {Liran}},
  \bibinfo {author} {\bibfnamefont {Y.}~\bibnamefont {Mazuz-Harpaz}}, \bibinfo
  {author} {\bibfnamefont {K.}~\bibnamefont {West}}, \bibinfo {author}
  {\bibfnamefont {L.}~\bibnamefont {Pfeiffer}},\ and\ \bibinfo {author}
  {\bibfnamefont {R.}~\bibnamefont {Rapaport}},\ }\href
  {https://doi.org/10.1126/sciadv.aat8880} {\bibfield  {journal} {\bibinfo
  {journal} {Science Advances}\ }\textbf {\bibinfo {volume} {4}},\ \bibinfo
  {pages} {eaat8880} (\bibinfo {year} {2018})},\ \Eprint
  {https://arxiv.org/abs/https://www.science.org/doi/pdf/10.1126/sciadv.aat8880}
  {https://www.science.org/doi/pdf/10.1126/sciadv.aat8880} \BibitemShut
  {NoStop}%
\bibitem [{\citenamefont {Tan}\ \emph {et~al.}(2020)\citenamefont {Tan},
  \citenamefont {Cotlet}, \citenamefont {Bergschneider}, \citenamefont
  {Schmidt}, \citenamefont {Back}, \citenamefont {Shimazaki}, \citenamefont
  {Kroner},\ and\ \citenamefont {\ifmmode \dot{I}\else
  \.{I}\fi{}mamo\ifmmode~\breve{g}\else \u{g}\fi{}lu}}]{Tan_2020}%
  \BibitemOpen
  \bibfield  {author} {\bibinfo {author} {\bibfnamefont {L.~B.}\ \bibnamefont
  {Tan}}, \bibinfo {author} {\bibfnamefont {O.}~\bibnamefont {Cotlet}},
  \bibinfo {author} {\bibfnamefont {A.}~\bibnamefont {Bergschneider}}, \bibinfo
  {author} {\bibfnamefont {R.}~\bibnamefont {Schmidt}}, \bibinfo {author}
  {\bibfnamefont {P.}~\bibnamefont {Back}}, \bibinfo {author} {\bibfnamefont
  {Y.}~\bibnamefont {Shimazaki}}, \bibinfo {author} {\bibfnamefont
  {M.}~\bibnamefont {Kroner}},\ and\ \bibinfo {author} {\bibfnamefont
  {A.~m.~c.}\ \bibnamefont {\ifmmode \dot{I}\else
  \.{I}\fi{}mamo\ifmmode~\breve{g}\else \u{g}\fi{}lu}},\ }\href
  {https://doi.org/10.1103/PhysRevX.10.021011} {\bibfield  {journal} {\bibinfo
  {journal} {Phys. Rev. X}\ }\textbf {\bibinfo {volume} {10}},\ \bibinfo
  {pages} {021011} (\bibinfo {year} {2020})}\BibitemShut {NoStop}%
\bibitem [{\citenamefont {Munoz-Matutano}\ \emph {et~al.}(2019)\citenamefont
  {Munoz-Matutano}, \citenamefont {Wood}, \citenamefont {Johnsson},
  \citenamefont {Vidal}, \citenamefont {Baragiola}, \citenamefont {Reinhard},
  \citenamefont {Lema\^itre}, \citenamefont {Bloch}, \citenamefont {Amo},
  \citenamefont {Nogues}, \citenamefont {Besga}, \citenamefont {Richard},\ and\
  \citenamefont {Volz}}]{Munoz_2019}%
  \BibitemOpen
  \bibfield  {author} {\bibinfo {author} {\bibfnamefont {G.}~\bibnamefont
  {Munoz-Matutano}}, \bibinfo {author} {\bibfnamefont {A.}~\bibnamefont
  {Wood}}, \bibinfo {author} {\bibfnamefont {M.}~\bibnamefont {Johnsson}},
  \bibinfo {author} {\bibfnamefont {X.}~\bibnamefont {Vidal}}, \bibinfo
  {author} {\bibfnamefont {B.~Q.}\ \bibnamefont {Baragiola}}, \bibinfo {author}
  {\bibfnamefont {A.}~\bibnamefont {Reinhard}}, \bibinfo {author}
  {\bibfnamefont {A.}~\bibnamefont {Lema\^itre}}, \bibinfo {author}
  {\bibfnamefont {J.}~\bibnamefont {Bloch}}, \bibinfo {author} {\bibfnamefont
  {A.}~\bibnamefont {Amo}}, \bibinfo {author} {\bibfnamefont {G.}~\bibnamefont
  {Nogues}}, \bibinfo {author} {\bibfnamefont {B.}~\bibnamefont {Besga}},
  \bibinfo {author} {\bibfnamefont {M.}~\bibnamefont {Richard}},\ and\ \bibinfo
  {author} {\bibfnamefont {T.}~\bibnamefont {Volz}},\ }\href
  {https://doi.org/https://doi.org/10.1038/s41563-019-0281-z} {\bibfield
  {journal} {\bibinfo  {journal} {Nature Materials}\ }\textbf {\bibinfo
  {volume} {18}},\ \bibinfo {pages} {213} (\bibinfo {year} {2019})}\BibitemShut
  {NoStop}%
\bibitem [{\citenamefont {Delteil}\ \emph {et~al.}(2019)\citenamefont
  {Delteil}, \citenamefont {Fink}, \citenamefont {Schade}, \citenamefont
  {H\"ofling}, \citenamefont {Schneider},\ and\ \citenamefont
  {Imamoglu}}]{Delteil_2019}%
  \BibitemOpen
  \bibfield  {author} {\bibinfo {author} {\bibfnamefont {A.}~\bibnamefont
  {Delteil}}, \bibinfo {author} {\bibfnamefont {T.}~\bibnamefont {Fink}},
  \bibinfo {author} {\bibfnamefont {A.}~\bibnamefont {Schade}}, \bibinfo
  {author} {\bibfnamefont {S.}~\bibnamefont {H\"ofling}}, \bibinfo {author}
  {\bibfnamefont {C.}~\bibnamefont {Schneider}},\ and\ \bibinfo {author}
  {\bibfnamefont {A.}~\bibnamefont {Imamoglu}},\ }\href@noop {} {\bibfield
  {journal} {\bibinfo  {journal} {Nature Materials}\ }\textbf {\bibinfo
  {volume} {18}},\ \bibinfo {pages} {219} (\bibinfo {year} {2019})}\BibitemShut
  {NoStop}%
\bibitem [{\citenamefont {Datta}\ \emph {et~al.}(2022)\citenamefont {Datta},
  \citenamefont {Khatoniar}, \citenamefont {Deshmukh}, \citenamefont {Thouin},
  \citenamefont {Bushati}, \citenamefont {De~Liberato}, \citenamefont {Cohen},\
  and\ \citenamefont {Menon}}]{Datta_2022}%
  \BibitemOpen
  \bibfield  {author} {\bibinfo {author} {\bibfnamefont {B.}~\bibnamefont
  {Datta}}, \bibinfo {author} {\bibfnamefont {M.}~\bibnamefont {Khatoniar}},
  \bibinfo {author} {\bibfnamefont {P.}~\bibnamefont {Deshmukh}}, \bibinfo
  {author} {\bibfnamefont {F.}~\bibnamefont {Thouin}}, \bibinfo {author}
  {\bibfnamefont {R.}~\bibnamefont {Bushati}}, \bibinfo {author} {\bibfnamefont
  {S.}~\bibnamefont {De~Liberato}}, \bibinfo {author} {\bibfnamefont {S.~K.}\
  \bibnamefont {Cohen}},\ and\ \bibinfo {author} {\bibfnamefont {V.~M.}\
  \bibnamefont {Menon}},\ }\href@noop {} {\bibfield  {journal} {\bibinfo
  {journal} {Nature Communications}\ }\textbf {\bibinfo {volume} {13}},\
  \bibinfo {pages} {6341} (\bibinfo {year} {2022})}\BibitemShut {NoStop}%
\bibitem [{\citenamefont {Inomata}\ \emph {et~al.}(2016)\citenamefont
  {Inomata}, \citenamefont {Lin}, \citenamefont {Koshino}, \citenamefont
  {Oliver}, \citenamefont {Tsai}, \citenamefont {Yamamoto},\ and\ \citenamefont
  {Nakamura}}]{Inomata_2016}%
  \BibitemOpen
  \bibfield  {author} {\bibinfo {author} {\bibfnamefont {K.}~\bibnamefont
  {Inomata}}, \bibinfo {author} {\bibfnamefont {Z.}~\bibnamefont {Lin}},
  \bibinfo {author} {\bibfnamefont {K.}~\bibnamefont {Koshino}}, \bibinfo
  {author} {\bibfnamefont {W.~D.}\ \bibnamefont {Oliver}}, \bibinfo {author}
  {\bibfnamefont {J.-S.}\ \bibnamefont {Tsai}}, \bibinfo {author}
  {\bibfnamefont {T.}~\bibnamefont {Yamamoto}},\ and\ \bibinfo {author}
  {\bibfnamefont {Y.}~\bibnamefont {Nakamura}},\ }\href
  {https://doi.org/10.1038/ncomms12303} {\bibfield  {journal} {\bibinfo
  {journal} {Nature Communications}\ }\textbf {\bibinfo {volume} {7}},\
  \bibinfo {pages} {12303} (\bibinfo {year} {2016})}\BibitemShut {NoStop}%
\bibitem [{\citenamefont {Lescanne}\ \emph {et~al.}(2020)\citenamefont
  {Lescanne}, \citenamefont {Del\'eglise}, \citenamefont {Albertinale},
  \citenamefont {R\'eglade}, \citenamefont {Capelle}, \citenamefont {Ivanov},
  \citenamefont {Jacqmin}, \citenamefont {Leghtas},\ and\ \citenamefont
  {Flurin}}]{Lescanne:2020}%
  \BibitemOpen
  \bibfield  {author} {\bibinfo {author} {\bibfnamefont {R.}~\bibnamefont
  {Lescanne}}, \bibinfo {author} {\bibfnamefont {S.}~\bibnamefont
  {Del\'eglise}}, \bibinfo {author} {\bibfnamefont {E.}~\bibnamefont
  {Albertinale}}, \bibinfo {author} {\bibfnamefont {U.}~\bibnamefont
  {R\'eglade}}, \bibinfo {author} {\bibfnamefont {T.}~\bibnamefont {Capelle}},
  \bibinfo {author} {\bibfnamefont {E.}~\bibnamefont {Ivanov}}, \bibinfo
  {author} {\bibfnamefont {T.}~\bibnamefont {Jacqmin}}, \bibinfo {author}
  {\bibfnamefont {Z.}~\bibnamefont {Leghtas}},\ and\ \bibinfo {author}
  {\bibfnamefont {E.}~\bibnamefont {Flurin}},\ }\href
  {https://doi.org/10.1103/PhysRevX.10.021038} {\bibfield  {journal} {\bibinfo
  {journal} {Phys. Rev. X}\ }\textbf {\bibinfo {volume} {10}},\ \bibinfo
  {pages} {021038} (\bibinfo {year} {2020})}\BibitemShut {NoStop}%
\bibitem [{\citenamefont {Besse}\ \emph {et~al.}(2018)\citenamefont {Besse},
  \citenamefont {Gasparinetti}, \citenamefont {Collodo}, \citenamefont
  {Walter}, \citenamefont {Kurpiers}, \citenamefont {Pechal}, \citenamefont
  {Eichler},\ and\ \citenamefont {Wallraff}}]{Besse:2018}%
  \BibitemOpen
  \bibfield  {author} {\bibinfo {author} {\bibfnamefont {J.-C.}\ \bibnamefont
  {Besse}}, \bibinfo {author} {\bibfnamefont {S.}~\bibnamefont {Gasparinetti}},
  \bibinfo {author} {\bibfnamefont {M.~C.}\ \bibnamefont {Collodo}}, \bibinfo
  {author} {\bibfnamefont {T.}~\bibnamefont {Walter}}, \bibinfo {author}
  {\bibfnamefont {P.}~\bibnamefont {Kurpiers}}, \bibinfo {author}
  {\bibfnamefont {M.}~\bibnamefont {Pechal}}, \bibinfo {author} {\bibfnamefont
  {C.}~\bibnamefont {Eichler}},\ and\ \bibinfo {author} {\bibfnamefont
  {A.}~\bibnamefont {Wallraff}},\ }\href
  {https://doi.org/10.1103/PhysRevX.8.021003} {\bibfield  {journal} {\bibinfo
  {journal} {Phys. Rev. X}\ }\textbf {\bibinfo {volume} {8}},\ \bibinfo {pages}
  {021003} (\bibinfo {year} {2018})}\BibitemShut {NoStop}%
\bibitem [{\citenamefont {Oppliger}\ \emph {et~al.}(2025)\citenamefont
  {Oppliger}, \citenamefont {Jang}, \citenamefont {Tarascio}, \citenamefont
  {Palma}, \citenamefont {Reichl}, \citenamefont {Wegscheider}, \citenamefont
  {Maisi}, \citenamefont {Zumb\"uhl},\ and\ \citenamefont
  {Scarlino}}]{Oppliger:2025}%
  \BibitemOpen
  \bibfield  {author} {\bibinfo {author} {\bibfnamefont {F.}~\bibnamefont
  {Oppliger}}, \bibinfo {author} {\bibfnamefont {W.}~\bibnamefont {Jang}},
  \bibinfo {author} {\bibfnamefont {A.}~\bibnamefont {Tarascio}}, \bibinfo
  {author} {\bibfnamefont {F.~D.}\ \bibnamefont {Palma}}, \bibinfo {author}
  {\bibfnamefont {C.}~\bibnamefont {Reichl}}, \bibinfo {author} {\bibfnamefont
  {W.}~\bibnamefont {Wegscheider}}, \bibinfo {author} {\bibfnamefont {V.~F.}\
  \bibnamefont {Maisi}}, \bibinfo {author} {\bibfnamefont {D.}~\bibnamefont
  {Zumb\"uhl}},\ and\ \bibinfo {author} {\bibfnamefont {P.}~\bibnamefont
  {Scarlino}},\ }\href {https://doi.org/10.48550/arXiv.2506.19828} {\bibinfo
  {title} {High-efficiency tunable microwave photon detector based on a
  semiconductor double quantum dot coupled to a superconducting high-impedance
  cavity}} (\bibinfo {year} {2025}),\ \Eprint
  {https://arxiv.org/abs/2506.19828} {arXiv:2506.19828} \BibitemShut {NoStop}%
\bibitem [{\citenamefont {Matern}\ \emph {et~al.}(2025)\citenamefont {Matern},
  \citenamefont {Biella}, \citenamefont {Scarlino}, \citenamefont {Carusotto},\
  and\ \citenamefont {Rastelli}}]{matern2025dispersive}%
  \BibitemOpen
  \bibfield  {author} {\bibinfo {author} {\bibfnamefont {S.}~\bibnamefont
  {Matern}}, \bibinfo {author} {\bibfnamefont {A.}~\bibnamefont {Biella}},
  \bibinfo {author} {\bibfnamefont {P.}~\bibnamefont {Scarlino}}, \bibinfo
  {author} {\bibfnamefont {I.}~\bibnamefont {Carusotto}},\ and\ \bibinfo
  {author} {\bibfnamefont {G.}~\bibnamefont {Rastelli}},\ }\href@noop {}
  {\bibfield  {journal} {\bibinfo  {journal} {arXiv preprint arXiv:2511.19128}\
  } (\bibinfo {year} {2025})}\BibitemShut {NoStop}%
\bibitem [{\citenamefont {Zeller}\ \emph {et~al.}(2026)\citenamefont {Zeller},
  \citenamefont {Danner}, \citenamefont {Hofheinz}, \citenamefont {Padurariu},
  \citenamefont {Ankerhold},\ and\ \citenamefont
  {Kubala}}]{zeller2026stroboscopic}%
  \BibitemOpen
  \bibfield  {author} {\bibinfo {author} {\bibfnamefont {H.}~\bibnamefont
  {Zeller}}, \bibinfo {author} {\bibfnamefont {L.}~\bibnamefont {Danner}},
  \bibinfo {author} {\bibfnamefont {M.}~\bibnamefont {Hofheinz}}, \bibinfo
  {author} {\bibfnamefont {C.}~\bibnamefont {Padurariu}}, \bibinfo {author}
  {\bibfnamefont {J.}~\bibnamefont {Ankerhold}},\ and\ \bibinfo {author}
  {\bibfnamefont {B.}~\bibnamefont {Kubala}},\ }\href
  {https://arxiv.org/abs/2603.16522} {\bibinfo {title} {Stroboscopic detection
  of itinerant microwave photons}} (\bibinfo {year} {2026}),\ \Eprint
  {https://arxiv.org/abs/2603.16522} {arXiv:2603.16522 [quant-ph]} \BibitemShut
  {NoStop}%
\bibitem [{\citenamefont {Blais}\ \emph {et~al.}(2021)\citenamefont {Blais},
  \citenamefont {Grimsmo}, \citenamefont {Girvin},\ and\ \citenamefont
  {Wallraff}}]{blais2021circuit}%
  \BibitemOpen
  \bibfield  {author} {\bibinfo {author} {\bibfnamefont {A.}~\bibnamefont
  {Blais}}, \bibinfo {author} {\bibfnamefont {A.~L.}\ \bibnamefont {Grimsmo}},
  \bibinfo {author} {\bibfnamefont {S.~M.}\ \bibnamefont {Girvin}},\ and\
  \bibinfo {author} {\bibfnamefont {A.}~\bibnamefont {Wallraff}},\ }\href@noop
  {} {\bibfield  {journal} {\bibinfo  {journal} {Reviews of Modern Physics}\
  }\textbf {\bibinfo {volume} {93}},\ \bibinfo {pages} {025005} (\bibinfo
  {year} {2021})}\BibitemShut {NoStop}%
\bibitem [{\citenamefont {Zenesini}\ \emph {et~al.}(2024)\citenamefont
  {Zenesini}, \citenamefont {Berti}, \citenamefont {Cominotti}, \citenamefont
  {Rogora}, \citenamefont {Moss}, \citenamefont {Billam}, \citenamefont
  {Carusotto}, \citenamefont {Lamporesi}, \citenamefont {Recati}, \citenamefont
  {Ferrari},\ and\ \citenamefont {et~al.}}]{Zenesini2024}%
  \BibitemOpen
  \bibfield  {author} {\bibinfo {author} {\bibfnamefont {A.}~\bibnamefont
  {Zenesini}}, \bibinfo {author} {\bibfnamefont {A.}~\bibnamefont {Berti}},
  \bibinfo {author} {\bibfnamefont {R.}~\bibnamefont {Cominotti}}, \bibinfo
  {author} {\bibfnamefont {C.}~\bibnamefont {Rogora}}, \bibinfo {author}
  {\bibfnamefont {I.~G.}\ \bibnamefont {Moss}}, \bibinfo {author}
  {\bibfnamefont {T.~P.}\ \bibnamefont {Billam}}, \bibinfo {author}
  {\bibfnamefont {I.}~\bibnamefont {Carusotto}}, \bibinfo {author}
  {\bibfnamefont {G.}~\bibnamefont {Lamporesi}}, \bibinfo {author}
  {\bibfnamefont {A.}~\bibnamefont {Recati}}, \bibinfo {author} {\bibfnamefont
  {G.}~\bibnamefont {Ferrari}},\ and\ \bibinfo {author} {\bibnamefont
  {et~al.}},\ }\href {https://doi.org/10.1038/s41567-023-02345-4} {\bibfield
  {journal} {\bibinfo  {journal} {Nature Physics}\ }\textbf {\bibinfo {volume}
  {20}},\ \bibinfo {pages} {558} (\bibinfo {year} {2024})}\BibitemShut
  {NoStop}%
\bibitem [{\citenamefont {Lagnese}\ \emph {et~al.}(2021)\citenamefont
  {Lagnese}, \citenamefont {Surace}, \citenamefont {Kormos},\ and\
  \citenamefont {Calabrese}}]{Lagnese:PRB2021}%
  \BibitemOpen
  \bibfield  {author} {\bibinfo {author} {\bibfnamefont {G.}~\bibnamefont
  {Lagnese}}, \bibinfo {author} {\bibfnamefont {F.~M.}\ \bibnamefont {Surace}},
  \bibinfo {author} {\bibfnamefont {M.}~\bibnamefont {Kormos}},\ and\ \bibinfo
  {author} {\bibfnamefont {P.}~\bibnamefont {Calabrese}},\ }\href
  {https://doi.org/10.1103/PhysRevB.104.L201106} {\bibfield  {journal}
  {\bibinfo  {journal} {Phys. Rev. B}\ }\textbf {\bibinfo {volume} {104}},\
  \bibinfo {pages} {L201106} (\bibinfo {year} {2021})}\BibitemShut {NoStop}%
\bibitem [{\citenamefont {Johansen}\ \emph {et~al.}(2025)\citenamefont
  {Johansen}, \citenamefont {Recati}, \citenamefont {Carusotto},\ and\
  \citenamefont {Biella}}]{Johansen:arXiv2025}%
  \BibitemOpen
  \bibfield  {author} {\bibinfo {author} {\bibfnamefont {C.}~\bibnamefont
  {Johansen}}, \bibinfo {author} {\bibfnamefont {A.}~\bibnamefont {Recati}},
  \bibinfo {author} {\bibfnamefont {I.}~\bibnamefont {Carusotto}},\ and\
  \bibinfo {author} {\bibfnamefont {A.}~\bibnamefont {Biella}},\ }\href@noop {}
  {\bibfield  {journal} {\bibinfo  {journal} {arXiv preprint arXiv:2508.13780}\
  } (\bibinfo {year} {2025})}\BibitemShut {NoStop}%
\bibitem [{\citenamefont {Maertens}\ \emph {et~al.}(2025)\citenamefont
  {Maertens}, \citenamefont {Haegeman},\ and\ \citenamefont
  {Van~Acoleyen}}]{Maertens:arXiv2025}%
  \BibitemOpen
  \bibfield  {author} {\bibinfo {author} {\bibfnamefont {D.}~\bibnamefont
  {Maertens}}, \bibinfo {author} {\bibfnamefont {J.}~\bibnamefont {Haegeman}},\
  and\ \bibinfo {author} {\bibfnamefont {K.}~\bibnamefont {Van~Acoleyen}},\
  }\href@noop {} {\bibfield  {journal} {\bibinfo  {journal} {arXiv preprint
  arXiv:2508.13645}\ } (\bibinfo {year} {2025})}\BibitemShut {NoStop}%
\bibitem [{\citenamefont {Coleman}(1977)}]{Coleman1977}%
  \BibitemOpen
  \bibfield  {author} {\bibinfo {author} {\bibfnamefont {S.}~\bibnamefont
  {Coleman}},\ }\href {https://doi.org/10.1103/PhysRevD.15.2929} {\bibfield
  {journal} {\bibinfo  {journal} {Phys. Rev. D}\ }\textbf {\bibinfo {volume}
  {15}},\ \bibinfo {pages} {2929} (\bibinfo {year} {1977})}\BibitemShut
  {NoStop}%
\bibitem [{\citenamefont {Devoto}\ \emph {et~al.}(2022)\citenamefont {Devoto},
  \citenamefont {Devoto}, \citenamefont {Di~Luzio},\ and\ \citenamefont
  {Ridolfi}}]{Devoto_2022}%
  \BibitemOpen
  \bibfield  {author} {\bibinfo {author} {\bibfnamefont {F.}~\bibnamefont
  {Devoto}}, \bibinfo {author} {\bibfnamefont {S.}~\bibnamefont {Devoto}},
  \bibinfo {author} {\bibfnamefont {L.}~\bibnamefont {Di~Luzio}},\ and\
  \bibinfo {author} {\bibfnamefont {G.}~\bibnamefont {Ridolfi}},\ }\href
  {https://doi.org/10.1088/1361-6471/ac7f24} {\bibfield  {journal} {\bibinfo
  {journal} {J. Phys. G: Nucl. Part. Phys.}\ }\textbf {\bibinfo {volume}
  {49}},\ \bibinfo {pages} {103001} (\bibinfo {year} {2022})}\BibitemShut
  {NoStop}%
\end{thebibliography}

%

\cleardoublepage
\thispagestyle{empty}

\appendix

\section{Frequency dependence of the jump probability}
\label{app:ph_freq}

In Sec.\ref{sec:analysis}, we have seen how the efficiency of the injection process of the incident photon wavepacket into the cavity volume depends on the incident photon spectral width $\kappa_e$ and how this effect has a substantial effect on the jump probability $\bar{P}_j$. In this Appendix we discuss the dependence of the jump probability on the frequency $\omega_e$ of the incident photon wavepacket.

For weak to intermediate values of the non-linearity strength parameter $g/\kappa$, the dominant excitation mode of the coherently-driven cavity is well captured by the normal Bogoliubov mode around the initial state $m_L$, whose frequency $\omega_{BN}$ can be expressed from the system parameters as \cite{Drummond_1980,Carusotto_2013}
\begin{equation}
    \omega_{BN}=\omega_d-\sqrt{(\Delta\omega+2gn_L)^2-(gn_L)^2},
    \label{eq:Bogo_freq}
\end{equation}

where $\Delta\omega\equiv\omega_c-\omega_d<-\sqrt{3}\kappa$. As the incident photon has the largest probability to enter the cavity when its frequency is close to resonance with the normal Bogoliubov mode~\cite{Claude:PRL2022}, it is natural to expect that the jump probability $\bar{P}_j$ will also be maximum around this resonance point.

\begin{figure}[h!]
    \centering
    \includegraphics[width=0.9\columnwidth]{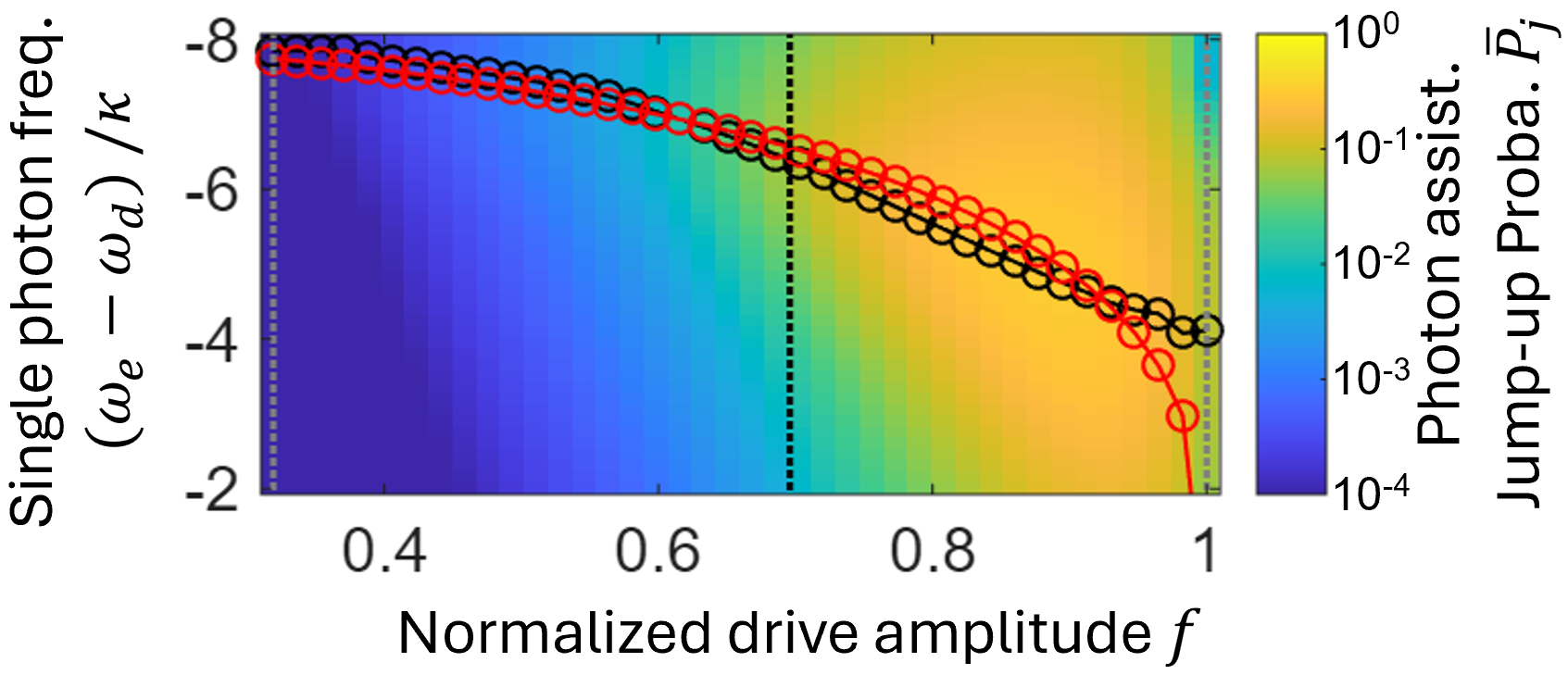}
    \caption{Color-plot (log-scale) of the single-photon-stimulated jump-up probability $\bar P_j$ as a function of the frequency $(\omega_e-\omega_d)/\kappa$ of the incident single photon and of the normalized drive amplitude $f=F_d/F_+$. The black and red symbols show the $f$-dependence of the largest jump-up probability point $\max_{\omega_e}(\bar P_j)$ of the normal Bogoliubov mode frequency $\omega_{BN}$. System parameters: $g=0.5\kappa$, $\kappa_e=0.3\kappa$ and $\Delta\omega=-8\kappa$.}
    \label{fig:fig8}
\end{figure}

To validate this statement, we calculated the jump-up probability $\bar{P}_{j}$ for a range of values of the coherent drive amplitude $F_d$ and of the incident photon frequency $\omega_e$. The result is shown as a color plot in Fig.\ref{fig:fig8}. In particular, a quantitative comparison is made between the frequency $\omega_{r,\max})$ at which the single-photon-stimulated jump-up probability $\bar P_{j}(\omega_e)$ is maximum (black), and the frequency $\omega_{BN}$ of the normal Bogoliubov mode (red). 

In spite of the quite large value of the nonlinearity $g/\kappa$ and the possibility of sizable corrections to the Bogoliubov theory, a very good agreement between the two curves is found for almost all operating points. Sizable deviations are only visible next to the right turning point, where the effect of the nonlinearity on the intra-cavity field dynamics is strongest and the jump dynamics is fastest.  

\section{Residence time and transit time for the incident photon}
\label{app:res_T}

The residence time $T_{res}$ of an incident photon in the cavity can be naturally defined as the time-integral of the probability for the photon to be in the cavity,
\begin{equation}
T_{res}=\int dt\, |a_{inc}(t)|^2
\end{equation}
where $a_{inc}(t)$ is the amplitude for the photon to be in the cavity at time $t$. This can be calculated in frequency space as
\begin{equation}
T_{res}=\int \frac{d\omega}{2\pi}\, |a_{inc}(\omega)|^2\,.
\label{eq:T_res_omega}
\end{equation}
Here, the cavity field can be expanded as
\begin{equation}
a_{inc}(\omega)=t_{cav}(\omega)\,\psi_{em}(\omega)
\end{equation}
where 
\begin{equation}
\psi_{em}(\omega)=\frac{i\sqrt{2\kappa_e}}{\omega_{em}-\omega-i\kappa_e}
\end{equation}
is the Fourier transform 
of the wavefunction 
\begin{equation}
\psi_{em}(t)=\sqrt{2\kappa_e}\Theta(t)\,e^{-i\omega_{em}t} e^{-\kappa_e t}
\end{equation}
of the incident photon emitted by the two-level system and the response of the cavity is assumed to be a Lorentzian
\begin{equation}
    t_{cav}(\omega)=\frac{\sqrt{2\kappa_{c,1}}}{\omega_{c}-\omega-i\kappa}
\end{equation}
centered at the cavity frequency $\omega_{c}$ and of width $\kappa$. 
Under our resonance condition $\omega_{inc}=\omega_{c}$, the integral \eqref{eq:T_res_omega} can be easily computed, giving the result 
\begin{equation}
T_{res}=\frac{2 \kappa_1}{\kappa(\kappa+\kappa_e)}\,.:
\label{eq:T_res_fin_bis}
\end{equation}
reported in \eqref{eq:T_res_fin}. While this expression is strictly valid only in the linear regime in the absence of coherent drive, it is a reasonable assumption to apply it throughout the lower branch of the bistability loop by just replacing the cavity frequency $\omega_c$ with the normal Bogoliubov mode frequency $\omega_{BN}$. Such an approximation is justified by the fact that the nonlinear frequency shift remains small $gn_L\ll|\Delta\omega|$ in this regime and the collective nature of the Bogoliubov mode has a minor impact.

Different from the residence time $T_{res}$, the average time $t_{tr}$ that a photon remains --once it is inside-- within the cavity can be estimated as the inverse half-width at half maximum $\kappa_{tr}$ of the cavity field spectrum $|a_{inc}(\omega)|^2$. At resonance, the inverse $\kappa_{tr}=1/t_{tr}$ of this transit time verifies 
\begin{equation}
    2\kappa_{tr}^2/\kappa^2=\sqrt{1+(\kappa_e/\kappa)^4+6(\kappa_e/\kappa)^2}-[1+(\kappa_e/\kappa)^2]\,.
\end{equation}
For short incident photon wavepackets $\kappa_e\gg\kappa$, the transit time is set by the cavity lifetime  $t_{tr}\simeq\kappa^{-1}$; for long wavepackets $\kappa_e\ll\kappa$, it is instead determined by the duration of the incident photon wavepacket $t_{tr}\simeq\kappa_e^{-1}$.

\section{Lower- and higher-state probabilities and occupancies during the dynamical evolution}
\label{app:Pj}

In this last Appendix, we provide the definition of the lower/higher state occupancies during the time evolution and of the jump probabilities. 

The single-photon-number distribution $p(n,t)$ for a state described by the density matrix $\rho(t)$ is given by
\begin{equation}
p(n,t)={\rm Tr} [\rho(t) |n\rangle \langle n|]\,.
\end{equation}
The probability of being in the lower/higher state can then be defined as 
\begin{equation}
P_{L(H)}(t)=\sum_{n \lessgtr \bar n} p(n,t)
\end{equation}
where the separation point $\bar n$ can be chosen close to the average of the mean-field values for lower and upper branch, $\bar n\simeq (n_{L,MF}+ n_{H,MF})/2$ in the steady-state. In general, the average photon number in the two states are defined as
\begin{equation}
n_{L(H)}(t)=\sum_{n \lessgtr \bar n} n \,  p(n,t)//P_{L(H)}\,.
\end{equation}

By employing the dynamical solution for the system density matrix $\rho^{(\alpha)}(t)$  with $\alpha=0,1$ corresponding to the case without or with incident photon respectively, we obtain the estimate for the jump enhancement factor in response to the incident photon as
\begin{equation}
{\cal F}(t)=P_H^{(1)}(t)/P_H^{(0)}(t)
\end{equation}
For each time, we define the jump probability $P_j$ from the relation
\begin{equation}
P_H^{(1)}(t)= P_H^{(0)}(t) + P_L^{(0)}(t) P_j(t),
\end{equation}
where the first term corresponds to probability of spontaneous jumps and the second term defines $P_j$ in terms of the conditional probability of jumping to the higher  branch given the probability of being in the lower branch in the absence of the incident photon. 
This yields  
\begin{equation}
P_j=\frac{P_H^{(1)}(t)-P_H^{(0)}(t)}{1-P_H^{(0)}(t)},
\end{equation}
from where we obtain the overall jump probability $\bar P_j= \max_t P_j(t)$ displayed in the main text.

The above quantities are related to the average populations in absence or presence of incident photon $n^{(0)}(t)$ and $n^{(1)}(t)$, respectively, according to:
\begin{equation}
n^\alpha(t)=P_{L}^\alpha(t) N_{L}^\alpha(t) +P_{H}^\alpha(t) N_{H}^\alpha(t)
\end{equation}
with $\alpha=(0)$, $(1)$.

\end{document}